\documentclass[notitlepage,nofootinbib,preprintnumbers,amssymb,superscriptaddress]{revtex4-1}
\usepackage{amsfonts,amssymb,amsmath,graphicx,color,bm}
\definecolor{ultramarine}{rgb}{0.07, 0.04, 0.56}
\definecolor{cadmiumgreen}{rgb}{0.0, 0.42, 0.24}
\definecolor{indigo(dye)}{rgb}{0.0, 0.25, 0.42}
\usepackage[linktocpage=true]{hyperref}
\hypersetup{
colorlinks=true,
citecolor=ultramarine,
linkcolor=cadmiumgreen,
urlcolor=indigo(dye),
}

\usepackage{autobreak}
\newcommand{\fr}[2]{\frac{#1}{#2}}
\newcommand{\pa}{\partial}
\newcommand{\ti}{\tilde}
\newcommand{\na}{\nabla}
\newcommand{\bra}[1]{\left( #1 \right)}  
\newcommand{\brb}[1]{\left[ #1 \right]}  
\newcommand{\brc}[1]{\left\{ #1 \right\}}  
\newcommand{\be}{\begin{equation}}  
\newcommand{\ee}{\end{equation}}
\newcommand{\bem}{\begin{bmatrix}}
\newcommand{\eem}{\end{bmatrix}}
\newcommand{\Mpl}{M_{\rm Pl}}

\newcommand{\ga}{\gamma}
\newcommand{\ep}{\epsilon}

\newcommand{\la}{\lambda}

\newcommand{\mn}{{\mu \nu}}
\newcommand{\mE}{\mathcal{E}}
\newcommand{\mJ}{\mathcal{J}}
\newcommand{\mF}{\mathcal{F}}
\newcommand{\mG}{\mathcal{G}}
\newcommand{\mH}{\mathcal{H}}
\newcommand{\mL}{\mathcal{L}}

\newcommand{\mS}{\mathcal{S}}

\allowdisplaybreaks

\begin{document}

\preprint{RUP-19-9, KOBE-COSMO-19-06, YITP-19-25}

\title{Linear stability analysis of hairy black holes in quadratic degenerate higher-order scalar-tensor theories: Odd-parity perturbations}

\author{Kazufumi Takahashi}
\affiliation{Department of Physics, Rikkyo University, Toshima, Tokyo 171-8501, Japan}
\affiliation{Department of Physics, Kobe University, Kobe 657-8501, Japan}

\author{Hayato Motohashi}
\affiliation{Center for Gravitational Physics, Yukawa Institute for Theoretical Physics, Kyoto University, Kyoto 606-8502, Japan}

\author{Masato Minamitsuji}
\affiliation{Centro de Astrof\'{\i}sica e Gravita\c c\~ao  - CENTRA, Departamento de F\'{\i}sica, Instituto Superior T\'ecnico - IST,
Universidade de Lisboa - UL, Av. Rovisco Pais 1, 1049-001 Lisboa, Portugal}

\begin{abstract}
We study static spherically symmetric black hole solutions with a linearly time-dependent scalar 
field and discuss their linear stability in the shift- and reflection-symmetric subclass of quadratic degenerate higher-order scalar-tensor (DHOST) theories. 
We present the explicit forms of the reduced system of background field equations for a generic theory within this subclass. 
Using the reduced equations of motion, we show that in several cases the solution is forced to be of the Schwarzschild or Schwarzschild--(anti-)de~Sitter form.
We consider odd-parity perturbations around general static spherically symmetric black hole solutions and derive the concise criteria for the black holes to be stable.
Our analysis also covers the case with a static or constant profile of the scalar field.
\end{abstract}

\maketitle

\section{Introduction}\label{sec:intro}

Modified gravity is a useful scheme for testing gravitation on 
cosmological scales and/or in the strong-field regimes.
There are several ways to relax the assumptions of the Lovelock's theorem~\cite{Lovelock:1971yv,Lovelock:1972vz}, and correspondingly several kinds of models of modified gravity,  
e.g., scalar-tensor theories, massive gravity, and higher-dimensional gravity~\cite{Berti:2015itd}. 
A common feature among them is that they have additional degree(s) of freedom (DOFs) on top of the metric.
Hence, scalar-tensor theories having only one additional DOF could help capturing fundamental aspects of such modified gravity theories with less technical complexity.

In the context of scalar-tensor theories, there have been a growing interest in a unified framework to incorporate the existing theories having higher derivatives in their Lagrangian (see Refs.~\cite{Langlois:2018dxi,Kobayashi:2019hrl} for recent reviews).
To this end, a crucial difficulty is the existence of unstable extra DOFs (known as ``Ostrogradsky ghosts'')~\cite{Woodard:2015zca} associated with higher-order derivatives in equations of motion (EOMs).
To circumvent this problem, one has to design a theory so that it allows a sufficient number of constraints to eliminate the extra DOFs in the Hamiltonian language, which is equivalent to require its Euler-Lagrange (EL) equations have degenerate higher-derivative terms~\cite{Motohashi:2014opa,Langlois:2015cwa,Motohashi:2016ftl,Klein:2016aiq,Motohashi:2017eya,Motohashi:2018pxg}.
The degeneracy allows one to reduce a priori higher-order EL equations to a system of lower-order differential equations.
A class of scalar-tensor theories that can trivially satisfy this requirement is the Horndeski theory~\cite{Horndeski:1974wa,Deffayet:2011gz,Kobayashi:2011nu}, which possesses the most general second-order EL equations for single-field scalar-tensor theories.
There are yet broader classes having degenerate higher-order EL equations, such as Gleyzes-Langlois-Piazza-Vernizzi~(GLPV, also known as ``beyond Horndeski'') theories~\cite{Gleyzes:2014dya} and degenerate higher-order scalar-tensor~(DHOST) theories~\cite{Langlois:2015cwa,Crisostomi:2016czh,BenAchour:2016fzp,Takahashi:2017pje,Langlois:2018jdg}.
DHOST theories provide the broadest class of scalar-tensor theories without Ostrogradsky ghost among those proposed so far respecting general covariance.\footnote{Yet, there have been attempts to further extend the framework by relaxing the requirement so that the degenerate property holds only in the unitary gauge~\cite{Gao:2014soa,DeFelice:2018mkq,Gao:2018znj}.}

The aim of the present paper is to investigate static spherically symmetric black hole~(BH) solutions in DHOST theories.
Analytic BH solutions in quadratic DHOST theories with a scalar field having a constant kinetic term were explored in Refs.~\cite{BenAchour:2018dap,Motohashi:2019sen,Charmousis:2019vnf}.
In Ref.~\cite{Motohashi:2019sen}, two of the authors of the present paper derived the spherically symmetric solutions with a linearly time-dependent scalar field in the shift-symmetric subclass of quadratic DHOST theories, 
which could reproduce the known solutions of the same type in the limit of the Horndeski~\cite{Babichev:2013cya,Kobayashi:2014eva} and GLPV~\cite{Babichev:2017lmw} theories. 
The stability of such BH solutions has been studied within the Horndeski theories~\cite{Ogawa:2015pea,Takahashi:2016dnv}, but a similar analysis for DHOST theories is still lacking.
Thus, the stability analysis presented in the present paper will be 
an extension and completion of the former studies. 
We shall consider the so-called ``class Ia'' (also called ``class ${}^2$N-I'') of quadratic DHOST theories~\cite{Langlois:2015cwa,Crisostomi:2016czh,Achour:2016rkg,BenAchour:2016fzp}, especially its shift- and reflection-symmetric subclass as it admits static BH solutions with a linearly time-dependent scalar field~\cite{Mukohyama:2005rw,Babichev:2013cya}.
This subclass allows ``stealth'' Schwarzschild metric (i.e., which is independent of the model parameters of the scalar sector, and thus the existence of the scalar field is hidden in the background spacetime geometry) and ``self-tuned'' Schwarzschild--(anti-)de~Sitter metric (i.e., whose effective cosmological constant is independent of the bare one) as an exact solution under certain conditions~\cite{Motohashi:2019sen}.
These solutions will also serve as a boundary condition
of analytic or numerical solutions in the presence of a compact object with a nontrivial matter profile in DHOST theories and thus be useful for future studies of their spacetime geometry and possible observational signatures.

It is important to note that one needs a special care for the stability analysis of BH solutions with a linearly time-dependent scalar field in the shift-symmetric scalar-tensor theories, since the quadratic action for the linear perturbations contains a nonvanishing cross term of time and spatial derivatives.
As pointed out in Refs.~\cite{Babichev:2017lmw,Babichev:2018uiw}, an unbounded Hamiltonian in a specific coordinate system does not necessarily mean instability of the system.
This is because a Hamiltonian is not a scalar quantity, and thus there may exist a coordinate system where the Hamiltonian is bounded below.
Indeed, as we will demonstrate in \S \ref{ssec:stability}, a coordinate transformation that eliminates the cross term in the quadratic action could make the Hamiltonian bounded below, even in the case where the Hamiltonian is unbounded in the original coordinate system.

The rest of this paper is organized as follows.
In \S \ref{sec:BG}, we define the model and study its static spherically symmetric BH solutions with a linearly time-dependent scalar field. 
We demonstrate how to 
reduce a priori higher-order EL equations to a lower-order system and investigate several specific cases where it is possible to obtain exact BH solutions.
In \S \ref{sec:pert}, we discuss the stability of BHs under linear odd-parity perturbations to obtain criteria for the BH solutions to be stable.
In \S \ref{sec:application}, we apply the stability criteria to several exact BH solutions.
Finally, we draw our conclusions in \S \ref{sec:conc}.

\section{Hairy black holes in shift- and reflection-symmetric quadratic DHOST theories}\label{sec:BG}

\subsection{The action}\label{ssec:model}

We consider the shift-symmetric subclass of quadratic DHOST theories~\cite{Langlois:2015cwa}, whose action has the form
	\be
	S=\int d^4x\sqrt{-g}\brb{F_0(X)+F_1(X)\Box\phi+F_2(X)R+\sum_{I=1}^{5}A_I(X)L_I^{(2)}}, \label{qDHOST}
	\ee
where $F_0$, $F_1$, $F_2$, and $A_I$ ($I=1,\cdots,5$) are functions of $X:=\phi_\mu\phi^\mu$ and 
	\be
	L_1^{(2)}:= \phi^{\mn}\phi_{\mn},~~~L_2^{(2)}:= (\Box\phi)^2,~~~L_3^{(2)}:= \phi^\mu\phi_{\mn}\phi^\nu\Box\phi,~~~L_4^{(2)}:= \phi^\mu\phi_{\mn}\phi^{\nu\la}\phi_\la,~~~L_5^{(2)}:= (\phi^\mu\phi_{\mn}\phi^\nu)^2,
	\ee 
with $\phi_\mu:= \na_\mu\phi$ and $\phi_{\mn}:= \na_\mu\na_\nu\phi$.
Clearly, the action respects the shift symmetry~$\phi\to \phi+{\rm const}$.
To avoid the generic problem of Ostrogradsky ghost associated with higher derivatives, one has to tune the coupling functions~$F_2$ and $A_I$'s,
so that the higher-derivative terms in the EL equations are degenerate.
All such possibilities have been exhausted in Ref.~\cite{Langlois:2015cwa} and further classified in Refs.~\cite{Crisostomi:2016czh,Achour:2016rkg}.
Among these classes is the ``class Ia''~\cite{Achour:2016rkg} (also known as ``class ${}^2$N-I''~\cite{BenAchour:2016fzp}), where $A_2$, $A_4$, and $A_5$ are written as
	\be
	\begin{split}
	A_2&=-A_1\ne -\fr{F_2}{X}, \\
	A_4&=\fr{1}{8(F_2-XA_1)^2}\bigl\{4F_2\brb{3(A_1-2F_{2X})^2-2A_3F_2}-A_3X^2(16A_1F_{2X}+A_3F_2) \\
	&~~~~~~~~~~~~~~~~~~~~~~~~+4X\bra{3A_1A_3F_2+16A_1^2F_{2X}-16A_1F_{2X}^2-4A_1^3+2A_3F_2F_{2X}}\bigr\}, \\
	A_5&=\fr{1}{8(F_2-XA_1)^2}(2A_1-XA_3-4F_{2X})\brb{A_1(2A_1+3XA_3-4F_{2X})-4A_3F_2},
	\end{split} \label{DC}
	\ee
while $F_0$, $F_1$, $F_2$, $A_1$, and $A_3$ remain arbitrary functions.
Here, a subscript~$X$ denotes a derivative with respect to $X$.
The condition~$A_1\ne F_2/X$ is necessary for the existence of two tensor modes~\cite{deRham:2016wji}.
This class includes Horndeski and GLPV theories up to quadratic-order interaction: The action reduces to that of GLPV theory by taking $A_1=2F_{2X}+\fr{X}{2}A_3$ and further to that of Horndeski theory by additionally imposing $A_3=0$.
It should also be noted that the other classes of quadratic DHOST theories are phenomenologically undesirable as they exhibit ghost/gradient instabilities in either of the tensor or scalar perturbations on a cosmological background, or otherwise the tensor DOFs do not propagate~\cite{Langlois:2017mxy,deRham:2016wji}.
Thus, throughout the present paper, we focus on the class Ia of quadratic DHOST theories.
Moreover, we restrict ourselves to the theories with $F_1=0$, i.e., those invariant under the reflection of the scalar field~$\phi\to-\phi$, which significantly simplifies the analysis.

Note that the shift- and reflection-symmetric subclass of the class Ia of quadratic DHOST theories is closed under disformal transformation~\cite{Bekenstein:1992pj} of the form,
	\be
	g_\mn \to \ti{g}_\mn:=\Omega(X)g_\mn+\Gamma(X)\phi_\mu\phi_\nu, \label{disformal}
	\ee
as long as $\Omega(\Omega-X\Omega_X-X^2\Gamma_X)\ne 0$, so that the transformation is invertible~\cite{Zumalacarregui:2013pma,Takahashi:2017zgr}.
This is consistent with the fact that an invertible transformation does not change the number of DOFs~\cite{Domenech:2015tca,Takahashi:2017zgr}.
The class Ia of quadratic DHOST theories is known to be recast into the Horndeski class via conformal/disformal transformation~\cite{Crisostomi:2016czh,Achour:2016rkg}.
Even though the structure of the Lagrangian would be simplified by such a transformation, in general the metric and scalar field profile would be modified,
while there exists a special case in which the disformal transformation preserves the form of solution and merely changes the parameters of the solution after a certain redefinition of coordinates (see Appendix~\ref{AppA}).

Before proceeding to the next section, let us comment on the constraints from the results of gravitational wave observations.
From the almost simultaneous detection of the gravitational waves~GW170817 and the $\gamma$-ray burst~170817A emitted from a binary neutron star merger~\cite{TheLIGOScientific:2017qsa,GBM:2017lvd,Monitor:2017mdv}, it turned out that the propagation speed of gravitational waves~($c_{\rm GW}$) coincides with that of light~($c_{\rm light}:=1$) to a precision of $10^{-15}$.
This can be satisfied for theories with $A_1=0$, where $c_{\rm GW}=1$ holds exactly for any cosmological background~\cite{Langlois:2017dyl}.
Another requirement is that the decay rate of gravitational waves into the scalar field should be small, which is satisfied if $A_3=0$~\cite{Creminelli:2018xsv}.
However, it should be remarked that these constraints apply only to the low-redshift universe ($z\lesssim 0.01$).
Furthermore, the energy scales observed by LIGO/Virgo lie close to the typical cutoff scale of many dark energy models~\cite{deRham:2018red}, so that it might be premature to exclude theories with $A_1\ne 0$ and/or $A_3\ne 0$. 
Hence, we leave $A_1$ and $A_3$ arbitrary unless otherwise stated.

\subsection{Reduction of background equations}\label{ssec:hairyBH}

Since we are interested in static spherically symmetric BH solutions, the background metric has the form,
	\be
	\bar{g}_{\mn}dx^\mu dx^\nu=-A(r)dt^2+\fr{dr^2}{B(r)}+2C(r)dtdr+D(r)r^2\ga_{ab}dx^adx^b, \label{sphsym}
	\ee
where $a$ and $b$ denote angular variables and $\ga_{ab}$ represents the metric on a two-dimensional sphere.
The scalar field is assumed to have a linearly time-dependent term as well as the $r$-dependent term,
	\be
	\phi(t,r)=qt+\psi(r), \label{lintimedep}
	\ee
where $q$ is a nonvanishing constant and
$\psi'\ne 0$, with a prime denoting differentiation with respect to $r$.
This is consistent with the static ansatz of the metric due to the shift symmetry of the theory~\cite{Mukohyama:2005rw,Babichev:2013cya}.
We shall discuss BH solutions with $q=0$ and/or $\psi'=0$ in \S \ref{ssec:qpsi=0}.
It should be noted that a rescaling of time coordinate~$t\to \alpha t$ amounts to the redefinition,
	\be
	A\to \alpha^2 A,~~~C\to \alpha C,~~~q\to \alpha q
	\ee
in Eqs.~\eqref{sphsym} and \eqref{lintimedep}, and the structure of the solution itself remains unchanged.

Substituting the ansatzes~\eqref{sphsym} and \eqref{lintimedep}, the action is written in terms of $A$, $B$, $C$, $D$, and $\psi$, from which we obtain the corresponding five EL equations~$\mE_\Phi=0$ ($\Phi=A,B,C,D,\psi$).
After deriving the EL equations, one may impose gauge fixing conditions~$C=0$ and $D=1$.
Note that not all of these equations are independent: 
$\mE_D=0$ and $\mE_\psi=0$ are automatically satisfied for any configuration that satisfies $\mE_A=0$, $\mE_B=0$, and $\mE_C=0$. 
Thus, in what follows, we study solutions of the EL equations for $A$, $B$, and $C$.

A crucial difference from the case of Horndeski theories is that each of the EL equations contains higher-derivative terms~$A''$ and $\psi'''$.
Nevertheless, thanks to the degeneracy of the theory, it is possible to arrange the system of EOMs into the one with at most second derivatives.
This can be achieved in a similar manner as in Ref.~\cite{Kobayashi:2018xvr}.
In what follows, we replace the derivatives of $\psi$ by $X$, $X'$, and $X''$ by use of
	\be
	X=-\fr{q^2}{A}+B\psi'{}^2. \label{ckt}
	\ee
First, by taking a linear combination of $\mE_B=0$ and $\mE_C=0$, one can simultaneously remove $A''$, $X''$, and $B'$, which allows us to express $B$ in terms of $(r,A,A',X,X')$,
	\begin{align}
	B^{-1}&=\fr{8(F_2-XA_1)+rX'(4F_{2X}-2A_1+XA_3)}{32A(r^2F_0+2F_2)(F_2-XA_1)^2}\bigl\{8(F_2-XA_1)\brb{(rA)'(F_2-XA_1)-q^2A_1} \nonumber \\
	&~~~+rX'\brb{2(2F_{2X}-A_1)\bra{3AF_2+(q^2-3XA)A_1}+(4q^2+3XA)A_3F_2-3(q^2+XA)XA_1A_3}\bigr\}. \label{solB}
	\end{align}
Next, substituting this into $\mE_A=0$ and $\mE_C=0$, we obtain equations of the form,
	\begin{align}
	K_1A''+K_2X''+J_1&=0, \label{EA2} \\
	K_1A''+K_2X''+J_2&=0, \label{EC2}
	\end{align}
with $K_1$, $K_2$, $J_1$, and $J_2$ being functions of $(r,A,A',X,X')$.
It is notable that, by virtue of the degeneracy, we have the same coefficients~$K_1$ and $K_2$ in front of $A''$ and $X''$, respectively, and thus the terms with $A''$ and $X''$ can be eliminated simultaneously.
As a consequence, we obtain a lower-order equation~$J_1=J_2$, which reads
	\be
	J_1-J_2=k_1A'+k_2X'+k_3=0, \label{J1-J2}
	\ee
where $k_1$, $k_2$, and $k_3$ are functions of $(r,A,X)$ (see Appendix \ref{AppB} for their explicit forms).
This can be solved for $X'$ as
	\be
	X'=-\fr{k_1A'+k_3}{k_2}. \label{solX'}
	\ee
Then, this equation can be used to eliminate $X'$ and $X''$ from Eq.~\eqref{EC2}.
This manipulation also removes $A''$, which is again due to the degeneracy of the theory. 
Hence, the resultant equation is written in terms of $r$, $A$, $A'$, and $X$, from which we obtain $A'$ in the form~$A'=\Psi_{1}(r,A,X)$.
Combining this with Eqs.~\eqref{solB} and \eqref{solX'}, we finally obtain a reduced system of EOMs written only by lower-order derivatives,
	\be
	A'=\Psi_{1}(r,A,X),~~~B=\Psi_{2}(r,X)A,~~~X'=\Psi_{3}(r,X). \label{redEOM}
	\ee
The expressions of $\Psi_1$, $\Psi_2$, and $\Psi_3$ are presented in Appendix \ref{AppB}.
As expected, the EOMs for the Horndeski theory are reproduced by taking the limit $A_1\to 2F_{2X}$ with $A_3\to 0$ in Eq.~\eqref{redEOM}.
For given coupling functions~$F_0$, $F_2$, $A_1$, and $A_3$, one can first solve the differential equation~$X'=\Psi_3(r,X)$ to obtain $X=X(r)$, which completely determines the ratio between $A$ and $B$ from $B=\Psi_2(r,X(r)) A$.
Then, from $A'=\Psi_1(r,A,X(r))$, one obtains $A$ as a function of $r$.

Note in passing that there exist models for which the above reduction procedure does not apply.
As an explicit example, we consider the case with $A_3=-2(F_2A_1)_X/F_2$, 
which includes the model with $A_1=A_3=0$ (i.e., those with $c_{\rm GW}=1$ and without decay of gravitons) as a special case.
In this case, Eq.~\eqref{solX'} does make sense, but it leads to $B^{-1}=0$ once substituted into Eq.~\eqref{solB}.
This explains why the above procedure fails for this particular class of models.
However, this does not imply this class does not allow physically sensible solutions.
We note that
	\be
	k_1A'+k_2X'+k_3\propto 2F_{0X}F_2(F_2-XA_1)-F_0(4F_2F_{2X}-F_2A_1-3XF_{2X}A_1-XF_2A_{1X}),
	\ee
so that Eq.~\eqref{J1-J2} is satisfied if
    \be
    F_{0X}=\fr{4F_2F_{2X}-F_2A_1-3XF_{2X}A_1-XF_2A_{1X}}{2F_2(F_2-XA_1)}F_0. \label{J1=J2-A1A3}
    \ee
This equation can be satisfied algebraically if
$X=X_0={\rm const}$, where $X_0$ is a real root of Eq.~\eqref{J1=J2-A1A3} (if it exists).
Then, Eq.~\eqref{EC2} is simplified as
	\be
	r(2F_2+r^2F_0)A''+4F_2A'-2rF_0A=-\fr{2q^2F_0A_1}{F_2-X_0A_1}r, \label{EC2-A1A3}
	\ee
where $F_0$, $F_2$, and $A_1$ are evaluated at $X=X_0$.
On the other hand, if there exists a constant~$\Lambda$ such that
    \be
    F_0=-2\Lambda F_2^{3/2}\sqrt{F_2-XA_1},
    \ee
then Eq.~\eqref{J1=J2-A1A3} is identically satisfied.
Note that now only $F_2$ and $A_1$ remain arbitrary.
It should be remarked that, as we will see in \S \ref{ssec:stability}, a stable BH solution satisfies $F_2>0$ and $F_2-XA_1>0$.
For this specific case, a disformal transformation,
    \be
    g_\mn\to \ti{g}_\mn=\sqrt{F_2(F_2-XA_1)}\bra{g_\mn+\fr{A_1}{F_2-XA_1}\phi_\mu\phi_\nu}
    \ee
brings the action 
into the form,
    \be
    S=\int d^4x\sqrt{-\ti{g}}\bra{\ti{R}-2\Lambda},
    \ee
i.e., general relativity with a cosmological constant~$\Lambda$.
The transformation is invertible unless 
$F_2^{3/2}/\sqrt{F_2-XA_1}\propto X$.
Therefore, in contrast to generic DHOST theories, this subclass has only two physical DOFs, and the system of EL equations is equivalent to those in general relativity~\cite{Takahashi:2017zgr}.
This means that one cannot determine all $A$, $B$, and $X$ in this subclass: Any of them remains an arbitrary function of $r$.
Similarly, there would also be a subtlety in BH solutions in the cuscuton theory~\cite{Afshordi:2006ad} and its extension~\cite{Iyonaga:2018vnu} having only two propagating DOFs when $\phi_\mu$ is timelike.
The same remark may also apply to the cuscuta-Galileon~\cite{deRham:2016ged} whose kinetic term can be reduced to a total derivative in flat spacetime.
We leave these issues for future study.

\subsection{Black holes with $q=0$ and/or $\psi'=0$}\label{ssec:qpsi=0}

Although we investigated BH solutions with $q\psi'\ne 0$ in the previous section, those where the scalar field have a static profile~$\phi=\psi(r)$ (i.e., $q=0$) or a constant profile (i.e., $q=0$ and $\psi'=0$) have also been extensively studied in the literature.
This type of solutions needs a special attention since $\mE_C$ can be written as
    \be
    \mE_C=q\psi'\ti{\mE}_C(r,A,A',A'',B,B',X,X',X''),
    \ee
and thus $\mE_C=0$ becomes trivial if $q\psi'=0$.
In fact, what we used in deriving the reduced EOMs~\eqref{redEOM} is not $\mE_C$ itself but rather $\ti{\mE}_C$.
Since $\mE_D=0$ can always be reproduced from the other EL equations, we focus on $\mE_A=\mE_B=\mE_\psi=0$ in the present section.

\subsubsection{The case of $\psi'=0$}

Let us first consider solutions with $\psi'=0$, which include $\phi={\rm const}$ as a special case.
In this case, $X$ is no longer an independent variable since $X=-q^2/A$.
It should also be noted that $\mE_\psi=0$ becomes trivial, and thus the independent equations are $\mE_A=\mE_B=0$.
One can first solve $\mE_B=0$ to obtain $B$ as a function of $(r,A,A')$.
Then, substituting this into $\mE_A=0$ yields a second-order differential equation for $A$.
Thus, we obtain $A$ and $B$ as functions of $r$.

\subsubsection{The case of $q=0$ and $\psi'\ne 0$}

Next, let us consider solutions with $q=0$ and $\psi'\ne 0$.
In this case, one has to take into account $\mE_\psi=0$, which can be integrated to give
    \be
    \ti{\mE}_C=\fr{c}{A},
    \ee
where $c$ is an integration constant.
Note that $\ti{\mE}_C=0$ is reproduced if $c=0$.
Thus, one can remove higher derivatives from the system of differential equations~$\mE_A=\mE_B=\mE_\psi=0$ in a similar manner as in \S \ref{ssec:hairyBH}.

To sum up, in either case, one has to consider a different system of EOMs from the one in the $q\psi'\ne 0$ case.
Therefore, one should treat the solutions with $q\psi'=0$ separately.
Nevertheless, at the level of linear odd-parity perturbations, one can treat all the solutions in a unified manner (see \S \ref{ssec:multipole}).

\subsection{Exact black holes with linearly time-dependent scalar hair}\label{ssec:BHsols}

Let us consider exact BH solutions with $q\psi'\ne 0$.
In Ref.~\cite{Motohashi:2019sen}, Schwarzschild--(anti-)de~Sitter metric with linearly time-dependent scalar field with a constant kinetic term was considered in shift-symmetric quadratic DHOST theories, and specific theories that allow them as exact solutions were identified. 
Here, we consider the reduced background equations~\eqref{redEOM} without assuming a specific metric solution and show that in several cases the solution is forced to be of the Schwarzschild or Schwarzschild--(anti-)de~Sitter form.

\subsubsection{Exact solutions for the case with $F_0=0$}

First, in the case of $F_0=0$, Eq.~\eqref{redEOM} is drastically simplified,
	\be
	A'=-\fr{q^2+XA}{rX},~~~B=-\fr{X}{q^2}A,~~~X'=0. \label{redEOM-F0}
	\ee
This can be straightforwardly integrated to give the Schwarzschild solution with a constant kinetic term,
	\be
	A=B=1-\fr{\mu}{r},~~~X=-q^2,  \label{stealth}
	\ee
where $\mu$ is an integration constant and we have rescaled $t$ so that $A=B$.
The coupling functions satisfy $F_0=F_1=0$ and $A_1+A_2=0$, which is consistent with the condition for the ``case~1'' of Ref.~\cite{Motohashi:2019sen}, and also the condition for the existence of general relativity solution in the limit of constant scalar field~\cite{Motohashi:2018wdq}.
We emphasize that here we derived the Schwarzschild solution without assuming it a priori.
Namely, for generic shift- and reflection-symmetric class Ia quadratic DHOST theories with $F_0=F_1=0$, the solution is forced to be the Schwarzschild metric with $X$ being a fixed constant $-q^2$. 
This generalizes the result of Ref.~\cite{Kobayashi:2018xvr}.

\subsubsection{Exact solutions with $X={\rm const}$}

Next, let us focus on solutions with $X=X_0={\rm const}$.
It is still possible for this case to obtain an exact solution even without specifying the theory.
Since $X'=0$, $\Psi_3(r,X_0)$ in Eq.~\eqref{redEOM} should vanish, which implies
	\be
	F_{0X}=\fr{8F_{2X}-2A_1+4XA_{1X}+3XA_3}{4(F_2-XA_1)}F_0 \label{constX}
	\ee
at $X=X_0$ [see Eq.~\eqref{Psi3}].
Note that Eq.~\eqref{constX} reduces to Eq.~\eqref{J1=J2-A1A3} by taking $A_3=-2(F_2A_1)_X/F_2$.
Generically, Eq.~\eqref{constX} will fix the value of $X_0$ (as long as it has a real solution).
Yet, for models where Eq.~\eqref{constX} becomes trivial (e.g., those with $F_0=0$), $X_0$ remains an arbitrary constant.
In either case, the remaining two EOMs in Eq.~\eqref{redEOM} reduce to
	\be
	A'=-\fr{q^2+X_0A}{X_0r}-\fr{q^2rF_0(X_0)}{2X_0[F_2(X_0)-X_0A_1(X_0)]},~~~B=-\fr{X_0}{q^2}A,
	\ee
which yield
	\be
	B=1-\fr{\mu}{r}+\fr{F_0(X_0)}{6[F_2(X_0)-X_0A_1(X_0)]}r^2=-\fr{X_0}{q^2}A,
        \label{sch_ds}
	\ee
with $\mu$ being an integration constant.
Moreover, if $X_0<0$, one can set $A=B$ (and thus $X_0=-q^2$) by rescaling $t$, so that we have the Schwarzschild--(anti-)de~Sitter solution.
Taking Eq.~\eqref{constX} into account, this case is consistent with ``case~1-$\Lambda$'' of Ref.~\cite{Motohashi:2019sen}.

\subsubsection{Exact solutions for the case with $A_3=-2(F_2A_1)_X/F_2$}

Finally, let us consider the case of $A_3=-2(F_2A_1)_X/F_2$.
As mentioned in \S \ref{ssec:hairyBH}, in this case the reduced equations are given by Eqs.~\eqref{J1=J2-A1A3} and \eqref{EC2-A1A3} instead of Eq.~\eqref{redEOM}.
Let us focus on solutions with $X=X_0={\rm const}$, where $X_0$ is a real root of Eq.~\eqref{J1=J2-A1A3}.
We can then integrate Eq.~\eqref{EC2-A1A3} to obtain 
	\be
	A=1-\fr{\mu}{r}+\fr{F_0(X_0)[F_2(X_0)-(X_0+q^2)A_1(X_0)]}{6F_2(X_0)[F_2(X_0)-X_0A_1(X_0)]}r^2, \label{x0neqq2}
	\ee
with $\mu$ being an integration constant.
Here, the other integration constant has been absorbed into a rescaling of $t$.
Then, substituting this result back into Eq.~\eqref{solB}, we have 
    \be \label{deficit-sol}
    A=\la B,~~~\la:=\fr{F_2(X_0)-(X_0+q^2)A_1(X_0)}{F_2(X_0)}.
    \ee
Since $\la\ne 1$ in general, we have the Schwarzschild--(anti-)de~Sitter solution with a deficit solid angle.
However, if $A_1=0$ or $X_0=-q^2$, we have $\la=1$, and thus there is no solid angle deficit.
These results are a natural generalization of the Schwarzschild-type solution with a deficit solid angle with static scalar field derived in Ref.~\cite{Motohashi:2019sen}, whose deficit solid angle also vanishes for $A_1=0$ or $X_0=-q^2$.  
The limiting case~$A_1=0$ or $X_0=-q^2$ of the solution~\eqref{deficit-sol} is consistent with ``case~2-$\Lambda$'' of Ref.~\cite{Motohashi:2019sen},
whereas the particular case~$F_0(X_0)=F_{0X}(X_0)=0$ with the Schwarzschild solution corresponds to ``case~2'' of Ref.~\cite{Motohashi:2019sen}.

Thus far, we have obtained exact BH solutions with $X={\rm const}$. 
By choosing the coupling functions in the Lagrangian appropriately, one may find an exact solution with a nonconstant $X$ as in Ref.~\cite{Babichev:2017guv}, but this is beyond the scope of the present paper.
Note that stability analysis in \S \ref{sec:pert} applies to any exact solution since we shall use only the background equations but not an explicit background solution.

\section{Odd-parity perturbations}\label{sec:pert}

To study perturbations of a spherically symmetric BH, it is useful to separate the deviation of the metric from its background value, $h_{\mn}:= g_{\mn}-\bar{g}_{\mn}$, into the odd- and even-parity perturbations
as odd- and even-parity modes are completely decoupled from each other unless the Lagrangian contains parity violating terms 
leading to analysis of coupled equations~\cite{Motohashi:2011pw,Motohashi:2011ds}. 
Since the analysis of the even-parity perturbations is technically involved, we consider only the odd-parity perturbations in the present paper.
The odd-parity perturbations can be decomposed as follows~\cite{Regge:1957td}:
	\be
	\begin{split}
	h_{tt}&=h_{tr}=h_{rr}=0,\\
	h_{ta}&=\sum_{\ell,m}h_{0,\ell m}(t,r)E_a{}^b\bar{\na}_bY_{\ell m}(\theta,\varphi),\\
	h_{ra}&=\sum_{\ell,m}h_{1,\ell m}(t,r)E_a{}^b\bar{\na}_bY_{\ell m}(\theta,\varphi),\\
	h_{ab}&=\sum_{\ell,m}h_{2,\ell m}(t,r)E_{(a}{}^c\bar{\na}_{|c|}\bar{\na}_{b)}Y_{\ell m}(\theta,\varphi),
	\end{split} \label{harmexp}
	\ee
where $Y_{\ell m}$ is the spherical harmonics, $E_{ab}$ is the completely antisymmetric tensor defined on a two-dimensional sphere, and $\bar{\na}_a$ denotes the covariant derivative with respect to $\gamma_{ab}$.
Since modes with different $(\ell,m)$ evolve independently, we focus on a specific mode and omit the indices~$\ell$ and $m$ unless necessary.
Note that the odd-parity perturbations do not have the monopole~($\ell=0$) mode and $h_2$ is vanishing for the dipole~($\ell=1$) modes.
Note also that we do not take into account the perturbation of the scalar field as it belongs to the even-parity perturbations.

The expansion coefficients $h_0$, $h_1$, and $h_2$ are not all physical DOFs as there exists a gauge DOF corresponding to the general covariance.
A general infinitesimal transformation of coordinates for the odd modes can be written as
	\be
	x^a\rightarrow x^a+\ep^a,~~~\ep_a:=\sum_{\ell,m}\Xi_{\ell m}(t,r)E_a{}^b\bar{\na}_bY_{\ell m}(\theta,\varphi).
	\ee
Correspondingly, the coefficients $h_0$, $h_1$, and $h_2$ transform as
	\be
	h_0\rightarrow h_0-\dot{\Xi},~~~
	h_1\rightarrow h_1-\Xi'+\fr{2}{r}\Xi,~~~
	h_2\rightarrow h_2-2\Xi, \label{gaugetrnsf}
	\ee
where a dot denotes a derivative with respect to $t$.
Therefore, in the case of $\ell\ge 2$, one can choose $\Xi=h_2/2$ to redefine $h_2=0$, which is a complete gauge fixing and thus can be imposed at the action level~\cite{Motohashi:2016prk}.
For the dipole modes where $h_2$ is absent, this gauge function~$\Xi$ is used to cancel out another unphysical DOF (see \S \ref{ssec:dipole}).

In what follows, we investigate $\ell\ge 2$ and $\ell=1$ modes separately and discuss the stability of BH solutions with a linearly time-dependent scalar 
field.

\subsection{Odd-parity perturbations with $\ell\ge 2$}\label{ssec:multipole}

First, we consider higher multipoles with $\ell\ge2$.
One can set $m=0$ from the beginning since all the terms with the same multipole index~$\ell$ contributes equally by virtue of the spherical symmetry of the background.
Hence, it is more useful to expand the metric perturbations in terms of the Legendre polynomials instead of the spherical harmonics. 
Thus, in the subsequent analysis, $h_0$, $h_1$, and $h_2$ denote the coefficients of $P_\ell(\cos \theta)$.
After performing the integration over angular variables, the second-order action
	\be
	S^{(2)}=\int dtdr\mL^{(2)}
	\ee
takes the form
	\be
	\fr{2\ell+1}{2\pi}\mL^{(2)}=a_1h_0^2+a_2h_1^2+a_3\bra{\dot{h}_1^2-2h'_0\dot{h}_1+h'_0{}^2+\fr{4h_0\dot{h}_1}{r}}+a_4h_0h_1. \label{Lag1}
	\ee
The coefficients here can be written as
	\be
	\begin{split}
	a_1&=\fr{\ell(\ell+1)}{r^2}\brb{\fr{d}{dr}\bra{r\sqrt{\fr{B}{A}}\mH}+\fr{(\ell-1)(\ell+2)}{2\sqrt{AB}}\mF},\\
	a_2&=-\fr{(\ell-1)\ell(\ell+1)(\ell+2)}{2}\fr{\sqrt{AB}}{r^2}\mG,\\
	a_3&=\fr{\ell(\ell+1)}{2}\sqrt{\fr{B}{A}}\mH,\\
	a_4&=\fr{(\ell-1)\ell(\ell+1)(\ell+2)}{r^2}\sqrt{\fr{B}{A}}\mJ,
	\end{split} \label{acoeff}
	\ee
with $\mF$, $\mG$, $\mH$, and $\mJ$ defined by
	\be
	\mF:= 2\bra{F_2+\fr{q^2}{A}A_1},~~~
	\mG:= 2\brb{F_2-\bra{\fr{q^2}{A}+X}A_1},~~~
	\mH:= 2(F_2-XA_1),~~~
	\mJ:= -2q\psi'A_1. \label{fghj}
	\ee
Note that we have used the background EOMs~\eqref{redEOM} in simplifying the quadratic Lagrangian.\footnote{One can verify that the quadratic Lagrangian for the case with $A_1=A_3=0$ is reproduced by taking the corresponding limit in Eqs.~\eqref{Lag1}--\eqref{fghj}, though the reduced background EOMs~\eqref{redEOM} does not apply in this case.
It is also possible to show that the second-order action for the solutions with $q=0$ and/or $\psi'=0$ discussed in \S \ref{ssec:qpsi=0} takes exactly the same form as \eqref{Lag1}, but with $\mJ=0$.
}
It should also be noted that the above expressions do not depend on $F_0$ or $A_3$ explicitly.
The explicit form of the quadratic Lagrangian would remain unchanged even if we have a nonvanishing $F_1$ in our Lagrangian, as is suggested by the result of Ref.~\cite{Takahashi:2016dnv}.
The result of Ref.~\cite{Ogawa:2015pea} for the shift- and reflection-symmetric Horndeski theory is reproduced by taking $A_1=2F_{2X}$.

Since the structure of the Lagrangian completely coincides with the one studied in Refs.~\cite{Ogawa:2015pea,Takahashi:2016dnv}, the subsequent analysis proceeds in a parallel manner.
Integrating by parts, one can rewrite Eq.~\eqref{Lag1} as
	\be
	\fr{2\ell+1}{2\pi}\mL^{(2)}=\bra{a_1-\fr{2(ra_3)'}{r^2}}h_0^2+a_2h_1^2+a_3\bra{\dot{h}_1-h_0'+\fr{2}{r}h_0}^2+a_4h_0h_1. \label{Lag2}
	\ee
Then, we introduce an auxiliary variable $\chi$ to write \cite{DeFelice:2011ka}
	\be
	\fr{2\ell+1}{2\pi}\mL^{(2)}=\bra{a_1-\fr{2(ra_3)'}{r^2}}h_0^2+a_2h_1^2+a_3\brb{-\chi^2+2\chi\bra{\dot{h}_1-h_0'+\fr{2}{r}h_0}}+a_4h_0h_1. \label{Lag3}
	\ee
Note that the EOM for $\chi$ yields $\chi=\dot{h}_1-h_0'+\fr{2}{r}h_0$, and substituting it back into Eq.~\eqref{Lag3} recovers the original Lagrangian~\eqref{Lag2}.
Once written in the form~\eqref{Lag3}, $h_0$ and $h_1$ become auxiliary fields, so that their EOMs yield
	\be
	h_0=-\fr{2r^2a_3a_4\dot{\chi}+4ra_2\brb{r(a_3\chi)'+2a_3\chi}}{4a_2\brb{r^2a_1-2(ra_3)'}-r^2a_4{}^2},~~~
	h_1=\fr{4a_3\dot{\chi}\brb{r^2a_1-2(ra_3)'}+2ra_4\brb{r(a_3\chi)'+2a_3\chi}}{4a_2\brb{r^2a_1-2(ra_3)'}-r^2a_4{}^2}. \label{h0h1}
	\ee
Then, the resubstitution into Eq.~\eqref{Lag3} yields the following Lagrangian written in terms of $\chi$:
	\be
	\fr{2\ell+1}{2\pi}\mL^{(2)}=\fr{\ell(\ell+1)}{2(\ell-1)(\ell+2)}\sqrt{\fr{B}{A}}\brc{b_1\dot{\chi}^2-b_2\chi'{}^2+b_3\dot{\chi}\chi'-\brb{\ell(\ell+1)b_4+V}\chi^2}, \label{Lag4}
	\ee
where 
    \be
	b_1=\fr{r^2\mF\mH^2}{A\mF\mG+B\mJ^2},~~~
	b_2=\fr{r^2AB\mG\mH^2}{A\mF\mG+B\mJ^2},~~~
	b_3=\fr{2r^2B\mH^2\mJ}{A\mF\mG+B\mJ^2},~~~
	b_4=\mH, \label{bcoeff}
	\ee
and $V$ is written as
	\be
	V=r^2\mH\brb{b_2\sqrt{\fr{B}{A}}\bra{\fr{1}{r^2\mH}\sqrt{\fr{A}{B}}}'\,}'-2\mH.
	\ee
The cross term $b_3\dot{\chi}\chi'$ is the crucial difference from the case with 
$q=0$ and/or $\psi'=0$.
Indeed, $b_3\propto \mJ\propto q\psi'$, and hence the cross term does not exist if $q\psi'=0$.

Note that one cannot rewrite the Lagrangian~\eqref{Lag3} in the form~\eqref{Lag4} for $\ell=1$ since $a_2=a_4=0$ in this case, and thus the denominators of $h_0$ and $h_1$ in Eq.~\eqref{h0h1} vanish.
We shall address the dipole perturbation in \S \ref{ssec:dipole}.

\subsection{Dipole perturbation: $\ell=1$}\label{ssec:dipole}

Let us focus on the dipole perturbation with $\ell=1$.
As was shown in Refs.~\cite{Kobayashi:2012kh,Ogawa:2015pea}, the dipole perturbations are related with the slow rotation of a BH.
Since the quadratic Lagrangian has the same form as the one in Ref.~\cite{Ogawa:2015pea}, we can follow the same procedure to clarify the physical meaning of the dipole perturbations.

We start from the Lagrangian~\eqref{Lag2} with $\ell=1$.
Since the coefficients $a_i$ ($i=1,\cdots,4$) defined by Eq.~\eqref{acoeff} satisfy
	\be
	a_1=\fr{2(ra_3)'}{r^2},~~~a_2=a_4=0
	\ee
for $\ell=1$, Eq.~\eqref{Lag2} is simplified as
	\be
	\fr{3}{2\pi}\mL^{(2)}=a_3\bra{\dot{h}_1-h_0'+\fr{2}{r}h_0}^2. \label{Lagdipole}
	\ee
We eliminate $h_1$ by choosing the gauge function~$\Xi$ appropriately.
Note that, as can be read off from Eq.~\eqref{gaugetrnsf}, there still remains a gauge DOF such that $\Xi=c(t)r^2$.
Then, the EOMs derived from Eq.~\eqref{Lagdipole} are
	\be
	\begin{split}
	h''_0+\fr{a_3'}{a_3}h'_0-\fr{2(ra_3)'}{r^2a_3}h_0&=0,\\
	\dot{h}'_0-\fr{2}{r}\dot{h}_0&=0.
	\end{split}
	\ee
The general solution to this system of equations is written as
	\be
	h_0=\fr{3Jr^2}{4\pi}\int^r \fr{dr'}{r'^4a_3(r')} 
	=\fr{3Jr^2}{8\pi}\int^r \sqrt{\fr{A}{B}}\fr{dr'}{r'^4(F_2-XA_1)}, \label{soldipole}
	\ee
where $J$ is an integration constant.
Here, the ($t$-dependent) integration constant resulting from the integration on $r$ can be set to zero by use of the residual gauge DOF mentioned above.
The result of Ref.~\cite{Ogawa:2015pea} is reproduced by taking the Horndeski limit where $A_1\to 2F_{2X}$.
$h_0$ given by Eq.~\eqref{soldipole} represents the slow rotation of a BH, with $J$ corresponding to the angular momentum of the BH.
If $X=X_0={\rm const}$, then the solution is generically the Schwarzschild--(anti-)de~Sitter metric with $X_0=-q^2$ (see \S \ref{ssec:BHsols}), and thus we obtain
	\be
	h_0=-\fr{J}{8\pi \brb{F_2(-q^2)+q^2A_1(-q^2)}r}. \label{fdf}
	\ee
In the case of $F_2=\Mpl^2/2$ and $A_1=0$, Eq.~\eqref{fdf} coincides with the frame-dragging function of the Kerr metric expanded up to first order in the angular momentum.

\subsection{Stability of $\ell\ge 2$ modes}\label{ssec:stability}

In this section, we discuss the linear stability of odd-parity modes with $\ell\ge 2$.

First, let us consider solutions with $q=0$ and/or $\psi'=0$.
As we mentioned above, for such a scalar field profile, the cross term $b_3\dot{\chi}\chi'$ in the Lagrangian~\eqref{Lag4} does not exist,
	\be \label{Lag-q0}
	\mL^{(2)}\propto \ti{\mL}:=\fr{1}{2}\sqrt{\fr{B}{A}}\brc{b_1 \dot \chi^2-b_2\chi'^2-\brb{\ell(\ell+1)b_4+V}\chi^2},
	\ee
where we have omitted the $r$-independent overall factor.
From this $\ti{\mL}$, one can construct the Hamiltonian as
	\be \label{Ham-q0}
	H=\fr{1}{2}\int dr\sqrt{\fr{B}{A}}\brc{\fr{A}{b_1B}\pi^2+b_2\chi'^2+\brb{\ell(\ell+1)b_4+V}\chi^2},
	\ee
with $\pi:=\pa \ti{\mL}/\pa \dot\chi$ being the canonical momentum conjugate to $\chi$.
Requiring $H$ should be bounded below, we obtain necessary conditions for the stability,
	\be
	b_1>0,~~~b_2>0,~~~b_4>0, \label{stabilityb-q0}
	\ee
namely, 
\be \mF>0,\quad \mG>0,\quad \mH>0, \label{stability1-q0} \ee
which is a natural generalization of the result for the Horndeski theory~\cite{Kobayashi:2012kh}.
By use of Eq.~\eqref{fghj}, these can further be translated into the following criteria on $F_2$ and $A_1$ evaluated at the background solution~$X=X(r)$:
	\be
	F_2>0,~~~F_2-XA_1>0. \label{stability2-q0}
	\ee
Interestingly, these conditions are consistent with the stability conditions for linear cosmological perturbations:~Regarding $F_2$ and $F_2-XA_1$ as those evaluated at the cosmological background, they guarantee that both the effective gravitational constant and the squared propagation speed of the tensor modes are positive~\cite{Langlois:2017mxy}.

Next, let us consider BH solutions with $q\psi'\ne 0$. 
With the nonvanishing cross term~$b_3\dot{\chi}\chi'$ in the Lagrangian~\eqref{Lag4}, one needs a special care.
As in the above discussion for the $q\psi'=0$ case, the (in)stability of a system is conveniently judged by the (un)boundedness of the Hamiltonian.
However, one should note that an unbounded Hamiltonian in a specific coordinate system does not necessarily mean that the system is unstable.
This is because a Hamiltonian is not a scalar quantity and thus there could be a coordinate transformation by which the originally unbounded Hamiltonian is mapped to a bounded one~\cite{Babichev:2017lmw,Babichev:2018uiw}.
To circumvent this subtlety, it is useful to perform a coordinate transformation such that the kinetic part of the quadratic Lagrangian~\eqref{Lag4} is diagonalized (i.e., the cross term~$\dot{\chi}\chi'$ vanishes).
This can be achieved by the following transformation of time coordinate~$t\to \ti{t}$,
	\be
	\ti{t}=t+\int \fr{b_3}{2b_2}dr. \label{diagonalize}
	\ee
In this new coordinate system, the Lagrangian is indeed diagonalized,
	\be
	\mL^{(2)}\propto \ti{\mL}:=\fr{1}{2}\sqrt{\fr{B}{A}}\brc{\ti{b}_1(\pa_{\ti{t}}\chi)^2-b_2\chi'^2-\brb{\ell(\ell+1)b_4+V}\chi^2},~~~
	\ti{b}_1:=b_1+\fr{b_3^2}{4b_2}=\fr{r^2\mH^2}{A\mG}, \label{diagLag}
	\ee
which should be compared with Eq.~\eqref{Lag-q0}.
Similar to Eq.~\eqref{Ham-q0}, we obtain the Hamiltonian as 
	\be
	H=\fr{1}{2}\int dr\sqrt{\fr{B}{A}}\brc{\fr{A}{\ti{b}_1B}\tilde\pi^2+b_2\chi'^2+\brb{\ell(\ell+1)b_4+V}\chi^2}, \label{Hamiltonian}
	\ee
where $\tilde\pi:=\pa \ti{\mL}/\pa (\pa_{\ti{t}}\chi)$.
Requiring $H$ should be bounded below, we obtain necessary conditions for the stability,
	\be
	\ti{b}_1>0,~~~b_2>0,~~~b_4>0, \label{stabilityb}
	\ee
which can be translated into the language of $(\mF,\mG,\mH,\mJ)$ as
	\be
	A\mF\mG+B\mJ^2>0,~~~\mG>0,~~~\mH>0. \label{stability1}
	\ee
Note that with $q=0$ or $\psi'=0$ we recover Eq.~\eqref{stability1-q0} as expected.
From Eq.~\eqref{fghj}, it is immediate to see $A\mF\mG+B\mJ^2=4AF_2(F_2-XA_1)=2AF_2\mH$.
Thus, we obtain the stability criteria, 
	\be
	F_2>0,~~~F_2-XA_1>0,~~~F_2-\bra{\fr{q^2}{A}+X}A_1>0, \label{stability4}
	\ee
where $F_2$, $A_1$, and $A$ are evaluated at the background solution.
Compared with the stability condition~\eqref{stability2-q0} for the static case, we obtain the third condition as a new condition.
Since $A\searrow0$ in the vicinity of the event horizon, the third condition can be satisfied only if $A_1\le 0$.\footnote{A possible loophole is that $A_1\searrow0$ and/or $F_2-XA_1\to +\infty$ near the horizon.}
Conversely, if $A_1\le 0$, the third condition is redundant as it follows from the second one.
Therefore, we obtain
    \be
	F_2>0,~~~F_2-XA_1>0,~~~A_1\le 0. \label{stability2}
	\ee
While \eqref{stability2} takes a simpler form, it should be regarded as a sufficient condition.  
Also, the stability criteria~\eqref{stability4} is disformally invariant (see Appendix~\ref{AppA}) and hence more fundamental.

Several remarks should be added here.
Unlike the discussion above, in the original coordinate system~$(t,r)$, the Hamiltonian constructed from the quadratic Lagrangian~\eqref{Lag4} is always unbounded below as long as $A_1\ne 0$ (see Refs.~\cite{Ogawa:2015pea,Takahashi:2015pad,Takahashi:2016dnv,Tretyakova:2017lyg} for related arguments).
Nevertheless, for those theories satisfying the conditions~\eqref{stability4}, the Hamiltonian calculated in the $(\ti{t},r)$~coordinate could be bounded below.
Another remark is that the conditions listed in Eq.~\eqref{stability1} ensure the positivity of the squared propagation speeds along the radial direction, $c_r^2$, and the angular direction, $c_\theta^2$, which are respectively given by
	\be
	c_r^2=\fr{A\mG^2}{A\mF\mG+B\mJ^2},~~~c_\theta^2=\fr{\mG}{\mH}.
	\ee
Note that, for theories with $A_1=0$, i.e., those with $c_{\rm GW}=1$ on a cosmological background, we have $\mF=\mG=\mH$ and $\mJ=0$, so that both $c_r$ and $c_\theta$ are unity.

Having known the necessary conditions for the linear stability, we further discuss the evolution of the perturbations to derive an additional condition for the stability.
We introduce the tortoise coordinate,
	\be
	r_\ast:=\int \fr{dr}{\sqrt{AB}} \label{tortoise}
	\ee
and use a new variable
	\be
	\Psi=\sqrt{\fr{b_2}{A}}\chi.
	\ee
In the remaining of this section, we regard $r$ as a function of $r_\ast$.
With these definitions, the EOM for $\Psi$ is written as
	\be
	\fr{\pa^2\Psi}{\pa r_\ast^2}-\fr{1}{c_r^2}\fr{\pa^2\Psi}{\pa \ti{t}^2}-V_{\rm eff}\Psi=0, \label{EOMPsi}
	\ee
where we have defined the effective potential,
	\be
	V_{\rm eff}:=\fr{AB}{b_2}\brb{\ell(\ell+1)\mH+V}+\sqrt{\fr{A}{b_2}}\fr{d^2}{dr_\ast^2}\bra{\sqrt{\fr{b_2}{A}}}. \label{Veff}
	\ee
For the Schwarzschild metric, $V_{\rm eff}$ coincides with the well-known Regge-Wheeler potential~\cite{Regge:1957td}.
Focusing on a mode with frequency~$\omega$, Eq.~\eqref{EOMPsi} takes the form of an eigenvalue equation,
	\be
	\hat{H}\Psi=\fr{\omega^2}{c_r^2}\Psi,~~~\hat{H}:=-\fr{d^2}{dr_\ast^2}+V_{\rm eff}.
	\ee
If all the eigenvalues of $\hat{H}$ are positive, the solution is stable for the fixed mode.
Note that $c_r^2>0$ as remarked above.
The positivity of the eigenvalues of $\hat{H}$ is equivalent to
	\be
	\langle\xi,\hat{H}\xi\rangle:=\int_E dr_\ast\bra{\left|\fr{d\xi}{dr_\ast}\right|^2+V_{\rm eff}|\xi|^2}>0, \label{positivity}
	\ee
where the integration runs over all the range of $r_\ast$ corresponding to the possible range of $r$ and $\xi(r_\ast)$ is an arbitrary square-integrable complex-valued function satisfying appropriate conditions at the boundary~$\pa E$.
Since the form of the effective potential~$V_{\rm eff}$ given by Eq.~\eqref{Veff} is involved and thus its positivity is far from trivial, we employ the ``$\mS$-deformation'' method~\cite{Kodama:2003jz,Ishibashi:2003ap,Kodama:2003kk} to {\it deform} the differential operator and the effective potential as
	\be
	\ti{D}:=\fr{d}{dr_\ast}+\mS,~~~
	\ti{V}:=V_{\rm eff}+\fr{d\mS}{dr_\ast}-\mS^2,
	\ee
with $\mS$ being an arbitrary real function of $r_\ast$.
Then, Eq.~\eqref{positivity} can be recast as
	\be
	\langle\xi,\hat{H}\xi\rangle=-\mS |\xi|^2\Bigr|_{\pa E}+\int_E dr_\ast\bra{\left|\ti{D}\xi\right|^2+\ti{V}|\xi|^2}>0.
	\ee
Hence, to complete the proof of the mode stability, it is enough to find a specific form of $\mS$ such that $\ti{V}>0$ for all $r$, as long as $\mS$ has a finite limit at the boundary.
This is indeed possible in the present case:~By choosing
	\be
	\mS=\fr{d}{dr_\ast}\brb{\ln \bra{r^2\mH\sqrt{\fr{B}{b_2}}}}=\fr{d}{dr_\ast}\brb{\ln \bra{r\sqrt{\fr{A\mF\mG+B\mJ^2}{A\mG}}}}, \label{S}
	\ee
we have
	\be
	\ti{V}=(\ell-1)(\ell+2)\fr{AB\mH}{b_2}>0.
	\ee
Thus, provided that the conditions in Eq.~\eqref{stability4} are satisfied, the odd-parity perturbations of BHs are fully stable for fixed modes as long as $\mS$ is finite at the boundary.

\section{Applications}\label{sec:application}

\subsection{Constant-scalar solution}

Let us apply the above criteria to solutions with $\phi=\phi_0={\rm const}$.
In this case, the stability condition~\eqref{stability2-q0} reads 
\be F_2(0)>0, \ee 
and the function~$\mS$ defined by Eq.~\eqref{S} is simplified as
    \be
    \mS=\fr{\sqrt{AB}}{r}.
    \ee

Now let us study the stability of the solutions found in Ref.~\cite{Motohashi:2019sen}.
If the condition 
\be F_0+2\Lambda F_2=0 \ee
is satisfied at $\phi=\phi_0$, the theory~\eqref{qDHOST} allows vacuum solutions in general relativity with a cosmological constant~$\Lambda$ as the exact solution~\cite{Motohashi:2018wdq}.
Specific examples are the Schwarzschild--(anti-)de~Sitter solutions for the two models~\cite{Motohashi:2019sen},
\be \label{csex1}
F_0 = - \Lambda \Mpl^2 + M^4 a(X), \quad
F_2 = \frac{\Mpl^2}{2} + M^2 b(X), 
\ee
where $a$ and $b$ are regular functions of $X$ satisfying $a(0)=b(0)=0$, and 
\be \label{csex2}
F_0 = - \Lambda [ \Mpl^2 + M^2 h(X) ], \quad
F_2 = \frac{\Mpl^2}{2} + \frac{M^2}{2} h(X), 
\ee
where $h$ is a regular function of $X$ and $h(0)\ne 0$ is allowed.
In either case, $\Lambda$ and $M$ are constant model parameters.
Note that $A_1$ and $A_3$ can be chosen freely so long as they are regular at $X=0$.
The first model~\eqref{csex1} identically satisfies the stability condition, whereas in the second model~\eqref{csex2}, the condition~$\Mpl^2+ M^2 h(0)>0$ is required for the stability of the solution with respect to the odd-parity perturbations.

\subsection{Constant-$X$ solution}

Let us proceed to solutions with $X=X_0={\rm const}$,
for which 
the stability conditions~\eqref{stability4} read
	\be
	F_2>0,~~~F_2-X_0A_1>0,~~~F_2-\bra{\fr{q^2}{A}+X_0}A_1>0,
	\ee
where $F_2$ and $A_1$ are evaluated at $X=X_0$,
and the function~$\mS$ defined by Eq.~\eqref{S} is simplified as
	\be
	\mS=\fr{\sqrt{AB}}{r}+\fr{q^2A_1A'\sqrt{AB}}{2A\brb{q^2A_1-A(F_2-X_0 A_1)}}.
	\label{S_x0}
	\ee
With the assumption~$X=X_0$, as we saw in \S \ref{ssec:BHsols}, we obtain the Schwarzschild--(anti-)de~Sitter metric with $X_0=-q^2$ 
without specifying the theory
[as long as Eq.~\eqref{constX} has a negative solution for $X$].
Therefore, 
the conditions~\eqref{stability4} read
	\be
	F_2>0,~~~F_2+q^2A_1>0,~~~F_2-\fr{1-A}{A}q^2A_1> 0, \label{stability3}
	\ee
where $F_2$ and $A_1$ are evaluated at $X=-q^2$.
Moreover, 
the function~$\mS$ in Eq.~\eqref{S_x0} is further simplified as
	\be
	\mS=\fr{A}{r}+\fr{q^2A_1A'}{2\brb{q^2A_1-A(F_2+q^2A_1)}}.
	\ee
Thus, if $A_1(-q^2)\ne 0$, then $\mS\to A'/2$ in the vicinity of the event and 
cosmological (in the case of the Schwarzschild-de~Sitter solutions) horizons where $A\to0$.
On the other hand, if $A_1(-q^2)=0$, then $\mS=0$ at both the horizons.
In either case, $\mS$ is finite at the event and cosmological horizons.
Moreover, for the Schwarzschild solution \eqref{stealth},
\be
\mS=
\fr{q^2(3r-2\mu)\mu A_1-2(r-\mu)^2F_2}
    {2r^2\brb{q^2\mu A_1-(r-\mu)F_2}},
\ee 
which decreases as $1/r$ and finite in the limit of $r\to \infty$.
Thus, the Schwarzschild and Schwarzschild-de~Sitter solutions
with a constant $X$ are stable at the level of linear odd-parity perturbations if the conditions~\eqref{stability3} are satisfied.

As specific examples, we consider the stealth Schwarzschild solution and Schwarzschild--(anti-)de~Sitter solutions for the class Ia quadratic DHOST theories where $A_1=0$ (i.e., $c_{\rm GW}=1$)
found in Ref.~\cite{Motohashi:2019sen}.
First, let us consider the model, 
\be F_0=M^4 a(X), \quad 
F_2= \frac{\Mpl^2}{2} + M^2 b(X), \quad
A_3= \frac{c(X)}{M^6},
\ee
where $a$, $b$, and $c$ are regular functions of $X$ and $M$ is a constant.
When $a(X)$ and $c(X)$ satisfy a certain set of conditions, this model allows the stealth Schwarzschild solution at $X=X_0={\rm const}$~\cite{Motohashi:2019sen}.
At this moment $b(X)$ remains a free function.
By requiring the odd-mode stability, we obtain an additional condition~$\Mpl^2 + 2M^2 b(X_0)>0$.
Second, we consider the model, 
\be 
F_0 = -\Mpl^2 \Lambda_{\rm b} + M^4 h(X), \quad
F_2 = \frac{\Mpl^2}{2} + \frac{\alpha}{2} M^2 h(X), \quad
A_3 = -8 \beta M^2 \frac{h_X(X)}{X},
\ee
where $h$ is a regular function of $X$ and $\Lambda_{\rm b}$, $M$, $\alpha$, and $\beta$ are constant model parameters.
The model allows various branches of self-tuned or untuned Schwarzschild--(anti-)de~Sitter solutions with $X=X_0={\rm const}$ depending on the form of the function~$h(X)$ and the values of dimensionless model parameters~$\alpha$ and $\beta$~\cite{Motohashi:2019sen}.
Again, by virtue of $A_1=0$, the odd-mode stability condition is simple: $\Mpl^2 + \alpha M^2 h(X_0)>0$, which applies to all the branches of solutions.

\section{Conclusions}\label{sec:conc}

We studied static spherically symmetric BH solutions with a linearly time-dependent scalar 
field in the class Ia of shift- and reflection-symmetric quadratic DHOST theories.
Although the background equations originally have higher-order derivatives, they can be reduced to a lower-order differential equation system~\eqref{redEOM} by taking their linear combinations, which was demonstrated in \S \ref{ssec:hairyBH}.
There are several specific situations where one can obtain exact BH solutions, e.g., if $F_0=0$ and/or $A_1=A_3=0$, or if we assume $X$ is constant.

We also analyzed the linear stability of the BH solutions under linear odd-parity perturbations in \S \ref{sec:pert}. 
In order for the BHs to be stable, it is necessary that the conditions listed in Eq.~\eqref{stability4} are satisfied. 
Furthermore, if these necessary conditions are satisfied and the function $\mS$ defined by Eq.~\eqref{S} is finite at the boundaries (i.e., the event/cosmological horizons and the spatial infinity),
then the BHs are fully stable at the level of linear odd-parity perturbations.
In \S \ref{sec:application}, we demonstrated an application of the stability criteria and the construction of $\mS$ to several exact BH solutions either with $\phi= {\rm const}$ or $X={\rm const}$, but we stress that the stability criteria and the construction of $\mS$ also apply to any solution where $\phi$ and/or $X$ is a nontrivial function of $r$. 
To complete the discussion on the linear stability, one has to study even-parity perturbations as well, which will be left for future study.


\acknowledgments{
K.T.\ was supported by JSPS Research Fellowships for Young Scientists No.\ 17J06778 and
JSPS KEKENHI Grant No.\ JP17H02894 and No.\ JP17K18778.
H.M.\ was supported by JSPS KEKENHI Grants No.\ JP17H06359 and No.\ JP18K13565.
M.M.\ was supported by the research grant under ``Norma Transit\'oria do DL 57/2016''.
}


\appendix

\section{BH solutions and their stability conditions in the disformal frame}\label{AppA}

In this appendix, we 
discuss BH solutions in another frame where the metric~$\ti{g}_\mn$ is related to $g_\mn$ in the original frame by the disformal transformation~\eqref{disformal}.
First, we summarize the transformation of the metric and scalar field, and show that their ansatzes are unchanged after an appropriate redefinition of coordinates.
As it should be, the background solution in the disformal frame is given by substituting the solution~$(g_\mn,\phi)$ in the original frame into Eq.~\eqref{disformal}.
The resultant metric is of the form,
	\be
	\ti{g}_{\mn}dx^\mu dx^\nu=-\bra{\Omega A-q^2\Gamma}dt^2+\fr{A\Omega+(q^2+AX)\Gamma}{AB}dr^2+2q\psi'\Gamma dtdr+r^2\Omega\ga_{ab}dx^adx^b.
	\ee
Performing a coordinate transformation~$t\to \tau=t-T(r)$ and $r\to \rho=r\sqrt{\Omega}$, where
	\be
	T'(r)=\fr{q\psi'\Gamma}{\Omega A-q^2\Gamma},
	\ee
the disformal metric becomes
	\be
	\ti{g}_{\mn}dx^\mu dx^\nu=-\ti{A}d\tau^2+\fr{d\rho^2}{\ti{B}}+\rho^2\ga_{ab}dx^adx^b, \label{disformalg}
	\ee
with
	\be
	\ti{A}:=\Omega A-q^2\Gamma,~~~
	\ti{B}:=\fr{B(\Omega A-q^2\Gamma)\rho'(r)^2}{A\Omega(\Omega+X\Gamma)}. \label{disformalA}
	\ee
Note that $r=r(\rho)$ is now a function of $\rho$ satisfying $r\sqrt{\Omega}=\rho$, and we require $\Omega A-q^2\Gamma>0$ so that $\ti{A}>0$.
In this new coordinate system, the scalar field is written as
    \be
    \phi=q\tau+\ti{\psi}(\rho),~~~\ti{\psi}(\rho):=\psi(r(\rho))+qT(r(\rho)),
    \ee
and thus linearly depends on $\tau$ with the same coefficient as in the original frame.
Furthermore, we have
    \be
    \ti{X}:=\ti{g}^\mn\ti{\na}_\mu\phi\ti{\na}_\nu\phi=\fr{X}{\Omega+X\Gamma}. \label{disformalX}
    \ee

Next, let us show the disformal invariance of the stability condition~\eqref{stability4}.
As remarked in \S \ref{ssec:model}, the shift- and reflection-symmetric subclass of the class Ia of quadratic DHOST theories is closed under the disformal transformation~\eqref{disformal}, so the disformal-frame Lagrangian is characterized by four functions~$\ti{F}_0$, $\ti{F}_2$, $\ti{A}_1$, and $\ti{A}_3$ of $\ti{X}$.
Following the same arguments as in the original frame, we obtain similar stability conditions for BHs as those in Eq.~\eqref{stability4},
	\be
	\ti{F}_2>0,~~~\ti{F}_2-\ti{X}\ti{A}_1>0,~~~\ti{F}_2-\bra{\fr{q^2}{\ti{A}}+\ti{X}}\ti{A}_1>0. \label{stability4-disformal}
	\ee
One would expect that the stability conditions in the disformal frame is equivalent to those in the original frame because of the invertibility of the disformal transformation~\eqref{disformal}.
This can be verified by rewriting the left-hand sides of the inequalities~\eqref{stability4-disformal} in terms of the original-frame quantities.
According to the results of Ref.~\cite{Achour:2016rkg} together with Eqs.~\eqref{disformalA} and \eqref{disformalX}, we have
	\be
	\begin{split}
	\ti{F}_2&=\fr{F_2}{\sqrt{\Omega(\Omega+X\Gamma)}}, \\
	\ti{F}_2-\ti{X}\ti{A}_1&=\fr{\sqrt{\Omega+X\Gamma}}{\Omega^{3/2}}(F_2-XA_1), \\
	\ti{F}_2-\bra{\fr{q^2}{\ti{A}}+\ti{X}}\ti{A}_1&=\sqrt{\fr{\Omega+X\Gamma}{\Omega}}\fr{A}{\Omega A-q^2\Gamma}\brb{F_2-\bra{\fr{q^2}{A}+X}A_1}.
	\end{split} \label{F2A1disformal}
	\ee
Hence, Eq.~\eqref{stability4-disformal} is reproduced from Eq.~\eqref{stability4}, and vice versa.

Finally, let us consider a specific solution and see how it (un)changes under the disformal transformation.
As we saw in \S \ref{ssec:BHsols}, once we impose $X=X_0={\rm const}$, the background solution is of the Schwarzschild--(anti-)de~Sitter form,
	\be
    A=1-\fr{\mu}{r}-\fr{\Lambda}{3}r^2=\la B.
	\ee
Here, for generality, we have introduced a parameter~$\la$, whose deviation from unity measures the solid angle deficit.
Note that $X_0\ne -q^2$ in general.
By rescaling $\tau$ as
	\be
	\tau\to \fr{\tau}{\sqrt{\Omega-q^2\Gamma}},
	\ee
the disformal metric in Eq.~\eqref{disformalg} is transformed into the form,
	\be
	\ti{g}_{\mn}dx^\mu dx^\nu\to -\bra{1-\fr{\ti{\mu}}{\rho}-\fr{\ti{\Lambda}}{3}\rho^2}d\tau^2+\fr{\ti{\la}}{1-\fr{\ti{\mu}}{\rho}-\fr{\ti{\Lambda}}{3}\rho^2}d\rho^2+\rho^2\ga_{ab}dx^adx^b,
	\ee
where we have defined $\ti{\mu}$, $\ti{\Lambda}$, and $\ti{\la}$ as
	\be
	\ti{\mu}:=\fr{\Omega^{3/2}\mu}{\Omega-q^2\Gamma},~~~
	\ti{\Lambda}:=\fr{\Lambda}{\Omega-q^2\Gamma},~~~
    \ti{\la}:=\fr{\Omega+X_0\Gamma}{\Omega-q^2\Gamma}\la.
\label{tilde_rel1}
	\ee
This is nothing but the Schwarzschild--(anti-)de~Sitter metric with a deficit solid angle.
Note that $\phi$ also gets transformed as
	\be
	\phi=Q\tau+\ti{\psi}(\rho),~~~Q:=\fr{q}{\sqrt{\Omega-q^2\Gamma}}.
\label{tilde_rel2}
	\ee
This means that the disformal transformation amounts to the redefinition of the parameters~$\mu$, $\Lambda$, $\la$, and $q$. 
Interestingly, even though the original-frame metric has no deficit solid angle (i.e., $\la=1$), the disformal-frame metric generically has a deficit solid angle if $X_0\ne -q^2$ and $\Gamma\ne 0$.
However, if $\la=1$ and $X_0=-q^2$, we have $\ti{\la}=1$ and $\ti{X}=-Q^2$.
An explicit Schwarzschild--(anti-)de~Sitter-type solution with a deficit solid angle was 
given in Eqs.~\eqref{x0neqq2} and \eqref{deficit-sol} under the condition,
\be
\la=\frac{F_2(X_0)-(X_0+q^2) A_1(X_0)}{F_2(X_0)}.
\ee
By use of Eqs.~\eqref{F2A1disformal}, \eqref{tilde_rel1}, and \eqref{tilde_rel2},
it is straightforward to confirm
\be
\ti{\la}=\frac{\ti{F}_2(\ti{X}_0)-(\ti{X}_0 +Q^2)\ti{A}_1(\ti{X}_0)}{\ti{F}_2(\ti{X}_0)},
\ee
where $\ti{X}_0:=X_0/[\Omega(X_0)+X_0\Gamma(X_0)]$.
Hence, the condition for the existence of the solution
with a deficit solid angle remains the same.

\section{Explicit forms of the reduced background EOMs}\label{AppB}

In this appendix, we present the explicit forms of functions which appeared in reducing the background EOMs (for their definitions, see \S \ref{ssec:hairyBH}).
$k_1$, $k_2$, and $k_3$ in Eq.~\eqref{J1-J2} are written as
	\begin{align}
	k_1&=8 r (F_{2} - X A_{1} )^2 [-4 r^2 F_{0X} F_{2} + 8 r^2 F_{0} F_{2X} + 8 X A_{1X} F_{2} + 
   4 X A_{3} F_{2} + 4 r^2 X A_{1X} F_{0} + 3 r^2 X A_{3} F_{0} \nonumber \\
	&~~~+  A_{1} (-2 r^2 F_{0} + 8 X F_{2X} + 4 r^2 X F_{0X})], \\
	k_2&=r \{4 A_{1}^3 (q^2 - 3 X A) [ r^2 F_{0} - 2 X (2 F_{2X} + r^2 F_{0X}) ] \nonumber \\
	&~~~+  F_{2} [48 r^2 A F_{2X} (-F_{0X} F_{2} + 2 F_{0} F_{2X}) + 
      4 A_{3} (4 F_{2} F_{2X} (q^2 + 3 X A) + 5 r^2 F_{0} F_{2X} (2 q^2 + 3 X A) - 
	  r^2 F_{0X} F_{2} (4 q^2 + 3 X A)) \nonumber \\  
	&~~~~~~~~~+  4 A_{1X} (2 F_{2} + r^2 F_{0} ) (4 F_{2X} (q^2 + 3 X A) + 
	X A_{3} (5 q^2 + 3 X A)) + 
	X A_{3}^2 (4 F_{2} (5 q^2 + 3 X A) + r^2 F_{0} (14 q^2 + 9 X A))] \nonumber \\ 
	&~~~+ 4 A_{1}^2 [q^2 (-2 r^2 F_{0} (4 F_{2X} + X A_{1X}) + X (2 F_{2X} + r^2 F_{0X} ) (4 F_{2X} - 3 X A_{3}) - 
         2 F_{2} (2 F_{2X} - r^2 F_{0X} + 2 X A_{1X} + X A_{3})) \nonumber \\
	&~~~~~~~~~+ 3 A (r^2 F_{0} (F_{2} + 2 X (3 F_{2X} + (A_{1X} + A_{3}) X)) \nonumber \\
	&~~~~~~~~~-  X (X (2 F_{2X} + r^2 F_{0X}) (4 F_{2X} + X A_{3}) + 
            F_{2} (4 F_{2X} + 4 r^2 F_{0X} - 2 (2 A_{1X} + A_{3}) X)))] \nonumber \\
	&~~~-A_{1} [3 X^2 A_{3}^2 (4 F_{2} + 3 r^2 F_{0}) (q^2 + X A) \nonumber \\ 
	&~~~~~~~~~+ 4 A_{1X} (2 F_{2} + r^2 F_{0}) (2 F_{2} (q^2 + 3 X A) + 
         X (-4 F_{2X} (q^2 - 3 X A) + 3 X A_{3} (q^2 + X A))) \nonumber \\
	&~~~~~~~~~+  4 A_{3} (F_{2} (2 X F_{2X} (-7 q^2 + 3 X A) + 2 F_{2} (q^2 + 3 X A) - 
            r^2 X F_{0X} (7 q^2 + 6 X A)) \nonumber \\
	&~~~~~~~~~+   r^2 F_{0} (3 F_{2} (q^2 + 2 X A) + X F_{2X} (2 q^2 + 15 X A))) \nonumber \\
	&~~~~~~~~~- 8 (2 q^2 F_{2X} (2 F_{2} F_{2X} - r^2 F_{0X} F_{2} + 3 r^2 F_{0} F_{2X} ) \nonumber \\
	&~~~~~~~~~+   3 A (r^2 F_{0X} F_{2} (F_{2} + 4 X F_{2X}) + 
            F_{2X} (4 X F_{2} F_{2X} - r^2 F_{0} (3 F_{2} + 4 X F_{2X}))))]\}, \\
	k_3&=8 (F_{2} - X A_{1} ) \{4 A_{1X} (2 F_{2} + r^2 F_{0}) (q^2 + X A) (F_{2} - X A_{1} ) \nonumber \\
	&~~~+ A_{3} [4 F_{2}^2 (q^2 + X A) - 4 X A_{1} F_{2} (q^2 + X A) 
	-  3 r^2 X A_{1} F_{0} (q^2 + X A) + r^2 F_{0} F_{2} (2 q^2 + 3 X A)] \nonumber \\
	&~~~+  2 [-2 r^2 A F_{2} (F_{0X} F_{2} - 2 F_{0} F_{2X}) 
	+   A_{1}^2 (q^2 + X A) (r^2 F_{0} - 2 (2 F_{2X} + r^2 F_{0X}) X) \nonumber \\
	&~~~~~~~~~+  A_{1} (4 F_{2} F_{2X} (q^2 + X A) - 2 r^2 F_{0} F_{2X} (q^2 + 2 X A) 
	+    r^2 F_{2} (-A F_{0} + 2 q^2 F_{0X} + 4 X A F_{0X} ))]\}.
	\end{align}
It is useful to introduce several auxiliary functions to write down the expressions for $\Psi_1$, $\Psi_2$, and $\Psi_3$ in Eq.~\eqref{redEOM}.
We write $\Psi_1$ as
	\be
	\fr{k_1}{8(F_2-XA_1)}\Psi_1=\fr{U}{V}-W,
	\ee
where $U$, $V$, and $W$ are defined as follows:
	\begin{align}
	U&=r^2 (2 A_{1X} F_{2} + A_{3} F_{2} + 2 A_{1} F_{2X}) (-2 A_{1} F_{0} - 4 F_{0X} F_{2} + 8 F_{0} F_{2X} + 
    4 X A_{1X} F_{0} + 3 X A_{3} F_{0} + 
    4 X A_{1} F_{0X}) \nonumber \\
	&~~~\times \{4 A_{1}^3 (q^2 - 3 X A) [r^2 F_{0} - 
       2 (2 F_{2X} + r^2 F_{0X}) X] \nonumber \\
	&~~~+ F_{2} [48 r^2 A F_{2X} (-F_{0X} F_{2} + 2 F_{0} F_{2X}) + 
       4 A_{3} (4 F_{2} F_{2X} (q^2 + 3 X A) + 5 r^2 F_{0} F_{2X} (2 q^2 + 3 X A) \nonumber \\
	&~~~~~~~~~-  r^2 F_{0X} F_{2} (4 q^2 + 3 X A)) + 
       4 A_{1X} (2 F_{2} + r^2 F_{0}) (4 F_{2X} (q^2 + 3 X A) + 
          X A_{3} (5 q^2 + 3 X A)) \nonumber \\
	&~~~~~~~~~+     X A_{3}^2 (4 F_{2} (5 q^2 + 3 X A) + r^2 F_{0} (14 q^2 + 9 X A))] \nonumber \\
	&~~~+ 4 A_{1}^2 [q^2 (-2 r^2 F_{0} (4 F_{2X} + X A_{1X}) + X (2 F_{2X} + r^2 F_{0X}) (4 F_{2X} - 3 X A_{3}) - 
          2 F_{2} (2 F_{2X} - r^2 F_{0X} + 2 X A_{1X} + X A_{3})) \nonumber \\
	&~~~~~~~~~+  3 A (r^2 F_{0} (F_{2} + 2 X (3 F_{2X} + (A_{1X} + A_{3}) X)) - 
          X (X (2 F_{2X} + r^2 F_{0X}) (4 F_{2X} + X A_{3}) \nonumber \\
	&~~~~~~~~~+   F_{2} (4 F_{2X} + 4 r^2 F_{0X} - 2 (2 A_{1X} + A_{3}) X)))] \nonumber \\
	&~~~- A_{1} [ 4 A_{1X} (2 F_{2} + r^2 F_{0}) (2 F_{2} (q^2 + 3 X A) + 
          X (-4 F_{2X} (q^2 - 3 X A) + 3 X A_{3} (q^2 + X A))) \nonumber \\
	&~~~~~~~~~+3 X^2 A_{3}^2 (4 F_{2} + 3 r^2 F_{0}) (q^2 + X A)+   
	 4 A_{3} (F_{2} (2 X F_{2X} (-7 q^2 + 3 X A) + 2 F_{2} (q^2 + 3 X A) - 
             r^2 X F_{0X} (7 q^2 + 6 X A)) \nonumber \\
	&~~~~~~~~~+    r^2 F_{0} (3 F_{2} (q^2 + 2 X A) + X F_{2X} (2 q^2 + 15 X A))) - 
       8 (2 q^2 F_{2X} (2 F_{2} F_{2X} - r^2 F_{0X} F_{2} + 3 r^2 F_{0} F_{2X} ) \nonumber \\
	&~~~~~~~~~+    3 A (r^2 F_{0X} F_{2} (F_{2} + 4 X F_{2X}) + 
             F_{2X} (4 X F_{2} F_{2X} - 
                r^2 F_{0} (3 F_{2} + 4 X F_{2X}))))]\}, \\
	V&=8 A_{1}^3 \{r^2 F_{0} (F_{2X} - 2 X F_{2XX} ) - 
    4 X [3 F_{2X}^2 - r^2 X F_{0X} F_{2XX} + r^2 F_{2X} (F_{0X} + X F_{0XX} )]\} \nonumber \\
	&~~~+ 4 A_{1}^2 \{16 F_{2} F_{2X}^2 +   8 F_{2} F_{2X} [r^2 F_{0X} + (-6 A_{1X} - 3 A_{3} + 2r^2 F_{0XX} ) X] \nonumber \\
	&~~~-  4 X F_{2X} [4 F_{2X}^2 + 2 r^2 X F_{0X} (2 A_{1X} + A_{3}) + 
       F_{2X} (4 r^2 F_{0X} + 8 X A_{1X} + 3 X A_{3})] \nonumber \\
	&~~~- 2 r^2 F_{0} [2 F_{2X}^2 - X^2 F_{2XX} (4 A_{1X} + 3 A_{3}) + 
       X F_{2X} (4 A_{1X} + 4 A_{3} + 4 X A_{1XX} + 3 X A_{3X} )] \nonumber \\
	&~~~+     r^2 F_{2} [4 F_{0} F_{2XX} - 4 X A_{1XX} F_{0} - 2 X A_{3X} F_{0} - 16 X F_{0X} F_{2XX} + 
       8 X^2 A_{1XX} F_{0X} + 4 X^2 A_{3X} F_{0X}  \nonumber \\
	&~~~~~~~~~+   2 A_{1X} (F_{0} - 4 X (F_{0X} + X F_{0XX} )) + 
       A_{3} (F_{0} - 4 X (F_{0X} + X F_{0XX} ))]\} \nonumber \\
	&~~~-  F_{2} \{-16 r^2 F_{2} (2 A_{1XX} + A_{3X}) (F_{0X} F_{2} - 2 F_{0} F_{2X}) + 
    64 X^2 A_{1X}^3 (2 F_{2} + r^2 F_{0}) + 3 X^2 A_{3}^3 (4 F_{2} + 3 r^2 F_{0}) \nonumber \\
	&~~~+     8 r^2 A_{3} [2 F_{0XX} F_{2}^2 - 6 F_{0X} F_{2} F_{2X} + 
       F_{0} (8 F_{2X}^2 - 4 F_{2} F_{2XX} + X A_{1XX} F_{2} )] \nonumber \\
	&~~~-  4 A_{3}^2 [3 r^2 F_{0} (F_{2} - 4 X F_{2X}) +    2 F_{2} (2 F_{2} - 2 X F_{2X} + 3 r^2 X F_{0X})] \nonumber \\
	&~~~- 8 A_{1X}^2 [2 F_{2} (4 F_{2} + X (-4 F_{2X} + 8 r^2 F_{0X} - 11 X A_{3})) 
	+    r^2 F_{0} (2 F_{2} - X (28 F_{2X} + 13 X A_{3}))] \nonumber \\
	&~~~+  2 A_{1X} [ X^2 A_{3}^2 (40 F_{2} + 27 r^2 F_{0}) - 
       4 r^2 (-4 F_{0XX} F_{2}^2 + 16 F_{0X} F_{2} F_{2X} +    F_{0} (-24 F_{2X}^2 + 8 F_{2} F_{2XX} + X A_{3X} F_{2} )) \nonumber \\
	&~~~~~~~~~-   4 A_{3} (r^2 F_{0} (4 F_{2} - 27 X F_{2X}) + 
          2 F_{2} (4 F_{2} - 4 X F_{2X} + 7 r^2 X F_{0X} ))]\} \nonumber \\
	&~~~+ 2A_{1} \{-16 X A_{1X}^2 F_{2} (3 F_{2} + 8 X F_{2X} + 4 r^2 X F_{0X}) - 
    3 X A_{3}^2 [4 F_{2}^2 + r^2 X F_{0} F_{2X} + 
       2 F_{2} (r^2 F_{0} + 4 X F_{2X} + 2 r^2 X F_{0X})] \nonumber \\
	&~~~+   4 A_{1X} [- r^2 X F_{0} F_{2X} (4 F_{2X} + X A_{3}) - 
       2 X F_{2} (2 F_{2X} + r^2 F_{0X}) (4 F_{2X} + 7 X A_{3}) \nonumber \\
	&~~~~~~~~~+  4 F_{2}^2 (4 F_{2X} + r^2 F_{0X} - 3 X A_{3} + 2 r^2 X F_{0XX}) + 
	   r^2 F_{0} F_{2} (4 F_{2X} - X (3 A_{3} + 12 F_{2XX} + X A_{3X} ))] \nonumber \\
	&~~~+ 4 r^2 [-4 F_{0XX} F_{2}^2 F_{2X} + 8 F_{0X} F_{2} F_{2X}^2 - 8 F_{0} F_{2X}^3 +    4 F_{0X} F_{2}^2 F_{2XX} + A_{3X} F_{2} (F_{0} F_{2} - 4 X F_{0X} F_{2} + 7 X F_{0} F_{2X}) \nonumber \\
	&~~~~~~~~~+  2 A_{1XX} F_{2} (-4 X F_{0X} F_{2} + F_{0} (F_{2} + 6 X F_{2X}))] \nonumber \\
	&~~~+    4 A_{3} [-5 r^2 X F_{0} F_{2X}^2 + 2 F_{2}^2 (4 F_{2X} + r^2 (F_{0X} + 2 X F_{0XX} )) \nonumber \\
	&~~~~~~~~~+  F_{2} (-2 X F_{2X} (4 F_{2X} + r^2 F_{0X}) + r^2 F_{0} (3 F_{2X} - 7 X F_{2XX}  + X^2 A_{1XX} ))]\}, \\
	W&=4 A_{1X} (2 F_{2} + r^2 F_{0}) (q^2 + X A) (F_{2} - X A_{1} ) \nonumber \\
	&~~~+  A_{3} [ 4F_{2}(q^2 + X A)(F_{2} - X A_{1})
	-3 r^2 X A_{1} F_{0} (q^2 + X A) + r^2 F_{0} F_{2} (2 q^2 + 3 X A)] \nonumber \\
	&~~~+ 2 \{-2 r^2 A F_{2} (F_{0X} F_{2} - 2 F_{0} F_{2X}) + 
    A_{1}^2 (q^2 + X A) [r^2 F_{0} - 2 X (2 F_{2X} + r^2 F_{0X}) ] \nonumber \\
	&~~~+   A_{1} [4 F_{2} F_{2X} (q^2 + X A) - 2 r^2 F_{0} F_{2X} (q^2 + 2 X A) + 
	r^2 F_{2} (-A F_{0} + 2 q^2 F_{0X} + 4 X A F_{0X} )]\}.
	\end{align}
With the above definitions, $\Psi_2$ and $\Psi_3$ are written as
	\begin{align}
	\Psi_2&=-\fr{2k_1}{q^2r(2F_2A_{1X}+F_2A_3+2F_{2X}A_1)[r(2A_1-XA_3-4F_{2X})\Psi_3-8(F_2-XA_1)]^2}, \label{Psi2} \\
	\Psi_3&=\fr{8r}{V}(F_2-XA_1)(2F_2A_{1X}+F_2A_3+2F_{2X}A_1)[F_0(2A_1-4XA_{1X}-3XA_3-8F_{2X})+4F_{0X}(F_2-XA_1)]. \label{Psi3}
	\end{align}


\bibliographystyle{mybibstyle}
\bibliography{BHinDHOST}

\begin{thebibliography}{63}%
\makeatletter
\providecommand \@ifxundefined [1]{%
 \@ifx{#1\undefined}
}%
\providecommand \@ifnum [1]{%
 \ifnum #1\expandafter \@firstoftwo
 \else \expandafter \@secondoftwo
 \fi
}%
\providecommand \@ifx [1]{%
 \ifx #1\expandafter \@firstoftwo
 \else \expandafter \@secondoftwo
 \fi
}%
\providecommand \natexlab [1]{#1}%
\providecommand \enquote  [1]{``#1''}%
\providecommand \bibnamefont  [1]{#1}%
\providecommand \bibfnamefont [1]{#1}%
\providecommand \citenamefont [1]{#1}%
\providecommand \href@noop [0]{\@secondoftwo}%
\providecommand \href [0]{\begingroup \@sanitize@url \@href}%
\providecommand \@href[1]{\@@startlink{#1}\@@href}%
\providecommand \@@href[1]{\endgroup#1\@@endlink}%
\providecommand \@sanitize@url [0]{\catcode `\\12\catcode `\$12\catcode
  `\&12\catcode `\#12\catcode `\^12\catcode `\_12\catcode `\%12\relax}%
\providecommand \@@startlink[1]{}%
\providecommand \@@endlink[0]{}%
\providecommand \url  [0]{\begingroup\@sanitize@url \@url }%
\providecommand \@url [1]{\endgroup\@href {#1}{\urlprefix }}%
\providecommand \urlprefix  [0]{URL }%
\providecommand \Eprint [0]{\href }%
\providecommand \doibase [0]{http://dx.doi.org/}%
\providecommand \selectlanguage [0]{\@gobble}%
\providecommand \bibinfo  [0]{\@secondoftwo}%
\providecommand \bibfield  [0]{\@secondoftwo}%
\providecommand \translation [1]{[#1]}%
\providecommand \BibitemOpen [0]{}%
\providecommand \bibitemStop [0]{}%
\providecommand \bibitemNoStop [0]{.\EOS\space}%
\providecommand \EOS [0]{\spacefactor3000\relax}%
\providecommand \BibitemShut  [1]{\csname bibitem#1\endcsname}%
\let\auto@bib@innerbib\@empty
\bibitem [{\citenamefont {Lovelock}(1971)}]{Lovelock:1971yv}%
  \BibitemOpen
  \bibfield  {author} {\bibinfo {author} {\bibfnamefont {D.}~\bibnamefont
  {Lovelock}},\ }\href {\doibase 10.1063/1.1665613} {\bibfield  {journal}
  {\bibinfo  {journal} {\emph {J. Math. Phys.}}\ }\textbf {\bibinfo {volume}
  {12}},\ \bibinfo {pages} {498} (\bibinfo {year} {1971})}\BibitemShut
  {NoStop}%
\bibitem [{\citenamefont {Lovelock}(1972)}]{Lovelock:1972vz}%
  \BibitemOpen
  \bibfield  {author} {\bibinfo {author} {\bibfnamefont {D.}~\bibnamefont
  {Lovelock}},\ }\href {\doibase 10.1063/1.1666069} {\bibfield  {journal}
  {\bibinfo  {journal} {\emph {J. Math. Phys.}}\ }\textbf {\bibinfo {volume}
  {13}},\ \bibinfo {pages} {874} (\bibinfo {year} {1972})}\BibitemShut
  {NoStop}%
\bibitem [{\citenamefont {Berti}\ \emph {et~al.}(2015)\citenamefont {Berti}
  \emph {et~al.}}]{Berti:2015itd}%
  \BibitemOpen
  \bibfield  {author} {\bibinfo {author} {\bibfnamefont {E.}~\bibnamefont
  {Berti}} \emph {et~al.},\ }\href {\doibase 10.1088/0264-9381/32/24/243001}
  {\bibfield  {journal} {\bibinfo  {journal} {\emph {Class. Quant. Grav.}}\
  }\textbf {\bibinfo {volume} {32}},\ \bibinfo {pages} {243001} (\bibinfo
  {year} {2015})},\ \Eprint {http://arxiv.org/abs/1501.07274} {arXiv:1501.07274
  [gr-qc]} \BibitemShut {NoStop}%
\bibitem [{\citenamefont {Langlois}(2019)}]{Langlois:2018dxi}%
  \BibitemOpen
  \bibfield  {author} {\bibinfo {author} {\bibfnamefont {D.}~\bibnamefont
  {Langlois}},\ }\href {\doibase 10.1142/S0218271819420069} {\bibfield
  {journal} {\bibinfo  {journal} {\emph {Int. J. Mod. Phys.}}\ }\textbf
  {\bibinfo {volume} {D28}},\ \bibinfo {pages} {1942006} (\bibinfo {year}
  {2019})},\ \Eprint {http://arxiv.org/abs/1811.06271} {arXiv:1811.06271
  [gr-qc]} \BibitemShut {NoStop}%
\bibitem [{\citenamefont {Kobayashi}(2019)}]{Kobayashi:2019hrl}%
  \BibitemOpen
  \bibfield  {author} {\bibinfo {author} {\bibfnamefont {T.}~\bibnamefont
  {Kobayashi}},\ }\href {\doibase 10.1088/1361-6633/ab2429} {\bibfield
  {journal} {\bibinfo  {journal} {\emph {Rept. Prog. Phys.}}\ }\textbf
  {\bibinfo {volume} {82}},\ \bibinfo {pages} {086901} (\bibinfo {year}
  {2019})},\ \Eprint {http://arxiv.org/abs/1901.07183} {arXiv:1901.07183
  [gr-qc]} \BibitemShut {NoStop}%
\bibitem [{\citenamefont {Woodard}(2015)}]{Woodard:2015zca}%
  \BibitemOpen
  \bibfield  {author} {\bibinfo {author} {\bibfnamefont {R.~P.}\ \bibnamefont
  {Woodard}},\ }\href {\doibase 10.4249/scholarpedia.32243} {\bibfield
  {journal} {\bibinfo  {journal} {\emph {Scholarpedia}}\ }\textbf {\bibinfo
  {volume} {10}},\ \bibinfo {pages} {32243} (\bibinfo {year} {2015})},\ \Eprint
  {http://arxiv.org/abs/1506.02210} {arXiv:1506.02210 [hep-th]} \BibitemShut
  {NoStop}%
\bibitem [{\citenamefont {Motohashi}\ and\ \citenamefont
  {Suyama}(2015)}]{Motohashi:2014opa}%
  \BibitemOpen
  \bibfield  {author} {\bibinfo {author} {\bibfnamefont {H.}~\bibnamefont
  {Motohashi}} and \bibinfo {author} {\bibfnamefont {T.}~\bibnamefont
  {Suyama}},\ }\href {\doibase 10.1103/PhysRevD.91.085009} {\bibfield
  {journal} {\bibinfo  {journal} {\emph {Phys. Rev.}}\ }\textbf {\bibinfo
  {volume} {D91}},\ \bibinfo {pages} {085009} (\bibinfo {year} {2015})},\
  \Eprint {http://arxiv.org/abs/1411.3721} {arXiv:1411.3721 [physics.class-ph]}
  \BibitemShut {NoStop}%
\bibitem [{\citenamefont {Langlois}\ and\ \citenamefont
  {Noui}(2016)}]{Langlois:2015cwa}%
  \BibitemOpen
  \bibfield  {author} {\bibinfo {author} {\bibfnamefont {D.}~\bibnamefont
  {Langlois}} and \bibinfo {author} {\bibfnamefont {K.}~\bibnamefont {Noui}},\
  }\href {\doibase 10.1088/1475-7516/2016/02/034} {\bibfield  {journal}
  {\bibinfo  {journal} {\emph {JCAP}}\ }\textbf {\bibinfo {volume} {1602}},\
  \bibinfo {pages} {034} (\bibinfo {year} {2016})},\ \Eprint
  {http://arxiv.org/abs/1510.06930} {arXiv:1510.06930 [gr-qc]} \BibitemShut
  {NoStop}%
\bibitem [{\citenamefont {Motohashi}\ \emph
  {et~al.}(2016{\natexlab{a}})\citenamefont {Motohashi}, \citenamefont {Noui},
  \citenamefont {Suyama}, \citenamefont {Yamaguchi},\ and\ \citenamefont
  {Langlois}}]{Motohashi:2016ftl}%
  \BibitemOpen
  \bibfield  {author} {\bibinfo {author} {\bibfnamefont {H.}~\bibnamefont
  {Motohashi}}, \bibinfo {author} {\bibfnamefont {K.}~\bibnamefont {Noui}},
  \bibinfo {author} {\bibfnamefont {T.}~\bibnamefont {Suyama}}, \bibinfo
  {author} {\bibfnamefont {M.}~\bibnamefont {Yamaguchi}},  and \bibinfo
  {author} {\bibfnamefont {D.}~\bibnamefont {Langlois}},\ }\href {\doibase
  10.1088/1475-7516/2016/07/033} {\bibfield  {journal} {\bibinfo  {journal}
  {\emph {JCAP}}\ }\textbf {\bibinfo {volume} {1607}},\ \bibinfo {pages} {033}
  (\bibinfo {year} {2016}{\natexlab{a}})},\ \Eprint
  {http://arxiv.org/abs/1603.09355} {arXiv:1603.09355 [hep-th]} \BibitemShut
  {NoStop}%
\bibitem [{\citenamefont {Klein}\ and\ \citenamefont
  {Roest}(2016)}]{Klein:2016aiq}%
  \BibitemOpen
  \bibfield  {author} {\bibinfo {author} {\bibfnamefont {R.}~\bibnamefont
  {Klein}} and \bibinfo {author} {\bibfnamefont {D.}~\bibnamefont {Roest}},\
  }\href {\doibase 10.1007/JHEP07(2016)130} {\bibfield  {journal} {\bibinfo
  {journal} {\emph {JHEP}}\ }\textbf {\bibinfo {volume} {07}},\ \bibinfo
  {pages} {130} (\bibinfo {year} {2016})},\ \Eprint
  {http://arxiv.org/abs/1604.01719} {arXiv:1604.01719 [hep-th]} \BibitemShut
  {NoStop}%
\bibitem [{\citenamefont {Motohashi}\ \emph
  {et~al.}(2018{\natexlab{a}})\citenamefont {Motohashi}, \citenamefont
  {Suyama},\ and\ \citenamefont {Yamaguchi}}]{Motohashi:2017eya}%
  \BibitemOpen
  \bibfield  {author} {\bibinfo {author} {\bibfnamefont {H.}~\bibnamefont
  {Motohashi}}, \bibinfo {author} {\bibfnamefont {T.}~\bibnamefont {Suyama}},
  and \bibinfo {author} {\bibfnamefont {M.}~\bibnamefont {Yamaguchi}},\ }\href
  {\doibase 10.7566/JPSJ.87.063401} {\bibfield  {journal} {\bibinfo  {journal}
  {\emph {J. Phys. Soc. Jap.}}\ }\textbf {\bibinfo {volume} {87}},\ \bibinfo
  {pages} {063401} (\bibinfo {year} {2018}{\natexlab{a}})},\ \Eprint
  {http://arxiv.org/abs/1711.08125} {arXiv:1711.08125 [hep-th]} \BibitemShut
  {NoStop}%
\bibitem [{\citenamefont {Motohashi}\ \emph
  {et~al.}(2018{\natexlab{b}})\citenamefont {Motohashi}, \citenamefont
  {Suyama},\ and\ \citenamefont {Yamaguchi}}]{Motohashi:2018pxg}%
  \BibitemOpen
  \bibfield  {author} {\bibinfo {author} {\bibfnamefont {H.}~\bibnamefont
  {Motohashi}}, \bibinfo {author} {\bibfnamefont {T.}~\bibnamefont {Suyama}},
  and \bibinfo {author} {\bibfnamefont {M.}~\bibnamefont {Yamaguchi}},\ }\href
  {\doibase 10.1007/JHEP06(2018)133} {\bibfield  {journal} {\bibinfo  {journal}
  {\emph {JHEP}}\ }\textbf {\bibinfo {volume} {06}},\ \bibinfo {pages} {133}
  (\bibinfo {year} {2018}{\natexlab{b}})},\ \Eprint
  {http://arxiv.org/abs/1804.07990} {arXiv:1804.07990 [hep-th]} \BibitemShut
  {NoStop}%
\bibitem [{\citenamefont {Horndeski}(1974)}]{Horndeski:1974wa}%
  \BibitemOpen
  \bibfield  {author} {\bibinfo {author} {\bibfnamefont {G.~W.}\ \bibnamefont
  {Horndeski}},\ }\href {\doibase 10.1007/BF01807638} {\bibfield  {journal}
  {\bibinfo  {journal} {\emph {Int. J. Theor. Phys.}}\ }\textbf {\bibinfo
  {volume} {10}},\ \bibinfo {pages} {363} (\bibinfo {year} {1974})}\BibitemShut
  {NoStop}%
\bibitem [{\citenamefont {Deffayet}\ \emph {et~al.}(2011)\citenamefont
  {Deffayet}, \citenamefont {Gao}, \citenamefont {Steer},\ and\ \citenamefont
  {Zahariade}}]{Deffayet:2011gz}%
  \BibitemOpen
  \bibfield  {author} {\bibinfo {author} {\bibfnamefont {C.}~\bibnamefont
  {Deffayet}}, \bibinfo {author} {\bibfnamefont {X.}~\bibnamefont {Gao}},
  \bibinfo {author} {\bibfnamefont {D.~A.}\ \bibnamefont {Steer}},  and
  \bibinfo {author} {\bibfnamefont {G.}~\bibnamefont {Zahariade}},\ }\href
  {\doibase 10.1103/PhysRevD.84.064039} {\bibfield  {journal} {\bibinfo
  {journal} {\emph {Phys. Rev.}}\ }\textbf {\bibinfo {volume} {D84}},\ \bibinfo
  {pages} {064039} (\bibinfo {year} {2011})},\ \Eprint
  {http://arxiv.org/abs/1103.3260} {arXiv:1103.3260 [hep-th]} \BibitemShut
  {NoStop}%
\bibitem [{\citenamefont {Kobayashi}\ \emph {et~al.}(2011)\citenamefont
  {Kobayashi}, \citenamefont {Yamaguchi},\ and\ \citenamefont
  {Yokoyama}}]{Kobayashi:2011nu}%
  \BibitemOpen
  \bibfield  {author} {\bibinfo {author} {\bibfnamefont {T.}~\bibnamefont
  {Kobayashi}}, \bibinfo {author} {\bibfnamefont {M.}~\bibnamefont
  {Yamaguchi}},  and \bibinfo {author} {\bibfnamefont {J.}~\bibnamefont
  {Yokoyama}},\ }\href {\doibase 10.1143/PTP.126.511} {\bibfield  {journal}
  {\bibinfo  {journal} {\emph {Prog. Theor. Phys.}}\ }\textbf {\bibinfo
  {volume} {126}},\ \bibinfo {pages} {511} (\bibinfo {year} {2011})},\ \Eprint
  {http://arxiv.org/abs/1105.5723} {arXiv:1105.5723 [hep-th]} \BibitemShut
  {NoStop}%
\bibitem [{\citenamefont {Gleyzes}\ \emph {et~al.}(2015)\citenamefont
  {Gleyzes}, \citenamefont {Langlois}, \citenamefont {Piazza},\ and\
  \citenamefont {Vernizzi}}]{Gleyzes:2014dya}%
  \BibitemOpen
  \bibfield  {author} {\bibinfo {author} {\bibfnamefont {J.}~\bibnamefont
  {Gleyzes}}, \bibinfo {author} {\bibfnamefont {D.}~\bibnamefont {Langlois}},
  \bibinfo {author} {\bibfnamefont {F.}~\bibnamefont {Piazza}},  and \bibinfo
  {author} {\bibfnamefont {F.}~\bibnamefont {Vernizzi}},\ }\href {\doibase
  10.1103/PhysRevLett.114.211101} {\bibfield  {journal} {\bibinfo  {journal}
  {\emph {Phys. Rev. Lett.}}\ }\textbf {\bibinfo {volume} {114}},\ \bibinfo
  {pages} {211101} (\bibinfo {year} {2015})},\ \Eprint
  {http://arxiv.org/abs/1404.6495} {arXiv:1404.6495 [hep-th]} \BibitemShut
  {NoStop}%
\bibitem [{\citenamefont {Crisostomi}\ \emph {et~al.}(2016)\citenamefont
  {Crisostomi}, \citenamefont {Koyama},\ and\ \citenamefont
  {Tasinato}}]{Crisostomi:2016czh}%
  \BibitemOpen
  \bibfield  {author} {\bibinfo {author} {\bibfnamefont {M.}~\bibnamefont
  {Crisostomi}}, \bibinfo {author} {\bibfnamefont {K.}~\bibnamefont {Koyama}},
  and \bibinfo {author} {\bibfnamefont {G.}~\bibnamefont {Tasinato}},\ }\href
  {\doibase 10.1088/1475-7516/2016/04/044} {\bibfield  {journal} {\bibinfo
  {journal} {\emph {JCAP}}\ }\textbf {\bibinfo {volume} {1604}},\ \bibinfo
  {pages} {044} (\bibinfo {year} {2016})},\ \Eprint
  {http://arxiv.org/abs/1602.03119} {arXiv:1602.03119 [hep-th]} \BibitemShut
  {NoStop}%
\bibitem [{\citenamefont {Ben~Achour}\ \emph
  {et~al.}(2016{\natexlab{a}})\citenamefont {Ben~Achour}, \citenamefont
  {Crisostomi}, \citenamefont {Koyama}, \citenamefont {Langlois}, \citenamefont
  {Noui},\ and\ \citenamefont {Tasinato}}]{BenAchour:2016fzp}%
  \BibitemOpen
  \bibfield  {author} {\bibinfo {author} {\bibfnamefont {J.}~\bibnamefont
  {Ben~Achour}}, \bibinfo {author} {\bibfnamefont {M.}~\bibnamefont
  {Crisostomi}}, \bibinfo {author} {\bibfnamefont {K.}~\bibnamefont {Koyama}},
  \bibinfo {author} {\bibfnamefont {D.}~\bibnamefont {Langlois}}, \bibinfo
  {author} {\bibfnamefont {K.}~\bibnamefont {Noui}},  and \bibinfo {author}
  {\bibfnamefont {G.}~\bibnamefont {Tasinato}},\ }\href {\doibase
  10.1007/JHEP12(2016)100} {\bibfield  {journal} {\bibinfo  {journal} {\emph
  {JHEP}}\ }\textbf {\bibinfo {volume} {12}},\ \bibinfo {pages} {100} (\bibinfo
  {year} {2016}{\natexlab{a}})},\ \Eprint {http://arxiv.org/abs/1608.08135}
  {arXiv:1608.08135 [hep-th]} \BibitemShut {NoStop}%
\bibitem [{\citenamefont {Takahashi}\ and\ \citenamefont
  {Kobayashi}(2017)}]{Takahashi:2017pje}%
  \BibitemOpen
  \bibfield  {author} {\bibinfo {author} {\bibfnamefont {K.}~\bibnamefont
  {Takahashi}} and \bibinfo {author} {\bibfnamefont {T.}~\bibnamefont
  {Kobayashi}},\ }\href {\doibase 10.1088/1475-7516/2017/11/038} {\bibfield
  {journal} {\bibinfo  {journal} {\emph {JCAP}}\ }\textbf {\bibinfo {volume}
  {1711}},\ \bibinfo {pages} {038} (\bibinfo {year} {2017})},\ \Eprint
  {http://arxiv.org/abs/1708.02951} {arXiv:1708.02951 [gr-qc]} \BibitemShut
  {NoStop}%
\bibitem [{\citenamefont {Langlois}\ \emph {et~al.}(2019)\citenamefont
  {Langlois}, \citenamefont {Mancarella}, \citenamefont {Noui},\ and\
  \citenamefont {Vernizzi}}]{Langlois:2018jdg}%
  \BibitemOpen
  \bibfield  {author} {\bibinfo {author} {\bibfnamefont {D.}~\bibnamefont
  {Langlois}}, \bibinfo {author} {\bibfnamefont {M.}~\bibnamefont
  {Mancarella}}, \bibinfo {author} {\bibfnamefont {K.}~\bibnamefont {Noui}},
  and \bibinfo {author} {\bibfnamefont {F.}~\bibnamefont {Vernizzi}},\ }\href
  {\doibase 10.1088/1475-7516/2019/02/036} {\bibfield  {journal} {\bibinfo
  {journal} {\emph {JCAP}}\ }\textbf {\bibinfo {volume} {1902}},\ \bibinfo
  {pages} {036} (\bibinfo {year} {2019})},\ \Eprint
  {http://arxiv.org/abs/1802.03394} {arXiv:1802.03394 [gr-qc]} \BibitemShut
  {NoStop}%
\bibitem [{\citenamefont {Gao}(2014)}]{Gao:2014soa}%
  \BibitemOpen
  \bibfield  {author} {\bibinfo {author} {\bibfnamefont {X.}~\bibnamefont
  {Gao}},\ }\href {\doibase 10.1103/PhysRevD.90.081501} {\bibfield  {journal}
  {\bibinfo  {journal} {\emph {Phys. Rev.}}\ }\textbf {\bibinfo {volume}
  {D90}},\ \bibinfo {pages} {081501} (\bibinfo {year} {2014})},\ \Eprint
  {http://arxiv.org/abs/1406.0822} {arXiv:1406.0822 [gr-qc]} \BibitemShut
  {NoStop}%
\bibitem [{\citenamefont {De~Felice}\ \emph {et~al.}(2018)\citenamefont
  {De~Felice}, \citenamefont {Langlois}, \citenamefont {Mukohyama},
  \citenamefont {Noui},\ and\ \citenamefont {Wang}}]{DeFelice:2018mkq}%
  \BibitemOpen
  \bibfield  {author} {\bibinfo {author} {\bibfnamefont {A.}~\bibnamefont
  {De~Felice}}, \bibinfo {author} {\bibfnamefont {D.}~\bibnamefont {Langlois}},
  \bibinfo {author} {\bibfnamefont {S.}~\bibnamefont {Mukohyama}}, \bibinfo
  {author} {\bibfnamefont {K.}~\bibnamefont {Noui}},  and \bibinfo {author}
  {\bibfnamefont {A.}~\bibnamefont {Wang}},\ }\href {\doibase
  10.1103/PhysRevD.98.084024} {\bibfield  {journal} {\bibinfo  {journal} {\emph
  {Phys. Rev.}}\ }\textbf {\bibinfo {volume} {D98}},\ \bibinfo {pages} {084024}
  (\bibinfo {year} {2018})},\ \Eprint {http://arxiv.org/abs/1803.06241}
  {arXiv:1803.06241 [hep-th]} \BibitemShut {NoStop}%
\bibitem [{\citenamefont {Gao}\ and\ \citenamefont {Yao}(2019)}]{Gao:2018znj}%
  \BibitemOpen
  \bibfield  {author} {\bibinfo {author} {\bibfnamefont {X.}~\bibnamefont
  {Gao}} and \bibinfo {author} {\bibfnamefont {Z.-B.}\ \bibnamefont {Yao}},\
  }\href {\doibase 10.1088/1475-7516/2019/05/024} {\bibfield  {journal}
  {\bibinfo  {journal} {\emph {JCAP}}\ }\textbf {\bibinfo {volume} {1905}},\
  \bibinfo {pages} {024} (\bibinfo {year} {2019})},\ \Eprint
  {http://arxiv.org/abs/1806.02811} {arXiv:1806.02811 [gr-qc]} \BibitemShut
  {NoStop}%
\bibitem [{\citenamefont {Ben~Achour}\ and\ \citenamefont
  {Liu}(2019)}]{BenAchour:2018dap}%
  \BibitemOpen
  \bibfield  {author} {\bibinfo {author} {\bibfnamefont {J.}~\bibnamefont
  {Ben~Achour}} and \bibinfo {author} {\bibfnamefont {H.}~\bibnamefont {Liu}},\
  }\href {\doibase 10.1103/PhysRevD.99.064042} {\bibfield  {journal} {\bibinfo
  {journal} {\emph {Phys. Rev.}}\ }\textbf {\bibinfo {volume} {D99}},\ \bibinfo
  {pages} {064042} (\bibinfo {year} {2019})},\ \Eprint
  {http://arxiv.org/abs/1811.05369} {arXiv:1811.05369 [gr-qc]} \BibitemShut
  {NoStop}%
\bibitem [{\citenamefont {Motohashi}\ and\ \citenamefont
  {Minamitsuji}(2019)}]{Motohashi:2019sen}%
  \BibitemOpen
  \bibfield  {author} {\bibinfo {author} {\bibfnamefont {H.}~\bibnamefont
  {Motohashi}} and \bibinfo {author} {\bibfnamefont {M.}~\bibnamefont
  {Minamitsuji}},\ }\href {\doibase 10.1103/PhysRevD.99.064040} {\bibfield
  {journal} {\bibinfo  {journal} {\emph {Phys. Rev.}}\ }\textbf {\bibinfo
  {volume} {D99}},\ \bibinfo {pages} {064040} (\bibinfo {year} {2019})},\
  \Eprint {http://arxiv.org/abs/1901.04658} {arXiv:1901.04658 [gr-qc]}
  \BibitemShut {NoStop}%
\bibitem [{\citenamefont {Charmousis}\ \emph {et~al.}(2019)\citenamefont
  {Charmousis}, \citenamefont {Crisostomi}, \citenamefont {Gregory},\ and\
  \citenamefont {Stergioulas}}]{Charmousis:2019vnf}%
  \BibitemOpen
  \bibfield  {author} {\bibinfo {author} {\bibfnamefont {C.}~\bibnamefont
  {Charmousis}}, \bibinfo {author} {\bibfnamefont {M.}~\bibnamefont
  {Crisostomi}}, \bibinfo {author} {\bibfnamefont {R.}~\bibnamefont {Gregory}},
   and \bibinfo {author} {\bibfnamefont {N.}~\bibnamefont {Stergioulas}},\
  }\Eprint {http://arxiv.org/abs/1903.05519} {arXiv:1903.05519 [hep-th]}
  \BibitemShut {NoStop}%
\bibitem [{\citenamefont {Babichev}\ and\ \citenamefont
  {Charmousis}(2014)}]{Babichev:2013cya}%
  \BibitemOpen
  \bibfield  {author} {\bibinfo {author} {\bibfnamefont {E.}~\bibnamefont
  {Babichev}} and \bibinfo {author} {\bibfnamefont {C.}~\bibnamefont
  {Charmousis}},\ }\href {\doibase 10.1007/JHEP08(2014)106} {\bibfield
  {journal} {\bibinfo  {journal} {\emph {JHEP}}\ }\textbf {\bibinfo {volume}
  {08}},\ \bibinfo {pages} {106} (\bibinfo {year} {2014})},\ \Eprint
  {http://arxiv.org/abs/1312.3204} {arXiv:1312.3204 [gr-qc]} \BibitemShut
  {NoStop}%
\bibitem [{\citenamefont {Kobayashi}\ and\ \citenamefont
  {Tanahashi}(2014)}]{Kobayashi:2014eva}%
  \BibitemOpen
  \bibfield  {author} {\bibinfo {author} {\bibfnamefont {T.}~\bibnamefont
  {Kobayashi}} and \bibinfo {author} {\bibfnamefont {N.}~\bibnamefont
  {Tanahashi}},\ }\href {\doibase 10.1093/ptep/ptu096} {\bibfield  {journal}
  {\bibinfo  {journal} {\emph {PTEP}}\ }\textbf {\bibinfo {volume} {2014}},\
  \bibinfo {pages} {073E02} (\bibinfo {year} {2014})},\ \Eprint
  {http://arxiv.org/abs/1403.4364} {arXiv:1403.4364 [gr-qc]} \BibitemShut
  {NoStop}%
\bibitem [{\citenamefont {Babichev}\ \emph
  {et~al.}(2018{\natexlab{a}})\citenamefont {Babichev}, \citenamefont
  {Charmousis}, \citenamefont {Esposito-Far\`ese},\ and\ \citenamefont
  {Leh\'ebel}}]{Babichev:2017lmw}%
  \BibitemOpen
  \bibfield  {author} {\bibinfo {author} {\bibfnamefont {E.}~\bibnamefont
  {Babichev}}, \bibinfo {author} {\bibfnamefont {C.}~\bibnamefont
  {Charmousis}}, \bibinfo {author} {\bibfnamefont {G.}~\bibnamefont
  {Esposito-Far\`ese}},  and \bibinfo {author} {\bibfnamefont {A.}~\bibnamefont
  {Leh\'ebel}},\ }\href {\doibase 10.1103/PhysRevLett.120.241101} {\bibfield
  {journal} {\bibinfo  {journal} {\emph {Phys. Rev. Lett.}}\ }\textbf {\bibinfo
  {volume} {120}},\ \bibinfo {pages} {241101} (\bibinfo {year}
  {2018}{\natexlab{a}})},\ \Eprint {http://arxiv.org/abs/1712.04398}
  {arXiv:1712.04398 [gr-qc]} \BibitemShut {NoStop}%
\bibitem [{\citenamefont {Ogawa}\ \emph {et~al.}(2016)\citenamefont {Ogawa},
  \citenamefont {Kobayashi},\ and\ \citenamefont {Suyama}}]{Ogawa:2015pea}%
  \BibitemOpen
  \bibfield  {author} {\bibinfo {author} {\bibfnamefont {H.}~\bibnamefont
  {Ogawa}}, \bibinfo {author} {\bibfnamefont {T.}~\bibnamefont {Kobayashi}},
  and \bibinfo {author} {\bibfnamefont {T.}~\bibnamefont {Suyama}},\ }\href
  {\doibase 10.1103/PhysRevD.93.064078} {\bibfield  {journal} {\bibinfo
  {journal} {\emph {Phys. Rev.}}\ }\textbf {\bibinfo {volume} {D93}},\ \bibinfo
  {pages} {064078} (\bibinfo {year} {2016})},\ \Eprint
  {http://arxiv.org/abs/1510.07400} {arXiv:1510.07400 [gr-qc]} \BibitemShut
  {NoStop}%
\bibitem [{\citenamefont {Takahashi}\ and\ \citenamefont
  {Suyama}(2017)}]{Takahashi:2016dnv}%
  \BibitemOpen
  \bibfield  {author} {\bibinfo {author} {\bibfnamefont {K.}~\bibnamefont
  {Takahashi}} and \bibinfo {author} {\bibfnamefont {T.}~\bibnamefont
  {Suyama}},\ }\href {\doibase 10.1103/PhysRevD.95.024034} {\bibfield
  {journal} {\bibinfo  {journal} {\emph {Phys. Rev.}}\ }\textbf {\bibinfo
  {volume} {D95}},\ \bibinfo {pages} {024034} (\bibinfo {year} {2017})},\
  \Eprint {http://arxiv.org/abs/1610.00432} {arXiv:1610.00432 [gr-qc]}
  \BibitemShut {NoStop}%
\bibitem [{\citenamefont {Ben~Achour}\ \emph
  {et~al.}(2016{\natexlab{b}})\citenamefont {Ben~Achour}, \citenamefont
  {Langlois},\ and\ \citenamefont {Noui}}]{Achour:2016rkg}%
  \BibitemOpen
  \bibfield  {author} {\bibinfo {author} {\bibfnamefont {J.}~\bibnamefont
  {Ben~Achour}}, \bibinfo {author} {\bibfnamefont {D.}~\bibnamefont
  {Langlois}},  and \bibinfo {author} {\bibfnamefont {K.}~\bibnamefont
  {Noui}},\ }\href {\doibase 10.1103/PhysRevD.93.124005} {\bibfield  {journal}
  {\bibinfo  {journal} {\emph {Phys. Rev.}}\ }\textbf {\bibinfo {volume}
  {D93}},\ \bibinfo {pages} {124005} (\bibinfo {year} {2016}{\natexlab{b}})},\
  \Eprint {http://arxiv.org/abs/1602.08398} {arXiv:1602.08398 [gr-qc]}
  \BibitemShut {NoStop}%
\bibitem [{\citenamefont {Mukohyama}(2005)}]{Mukohyama:2005rw}%
  \BibitemOpen
  \bibfield  {author} {\bibinfo {author} {\bibfnamefont {S.}~\bibnamefont
  {Mukohyama}},\ }\href {\doibase 10.1103/PhysRevD.71.104019} {\bibfield
  {journal} {\bibinfo  {journal} {\emph {Phys. Rev.}}\ }\textbf {\bibinfo
  {volume} {D71}},\ \bibinfo {pages} {104019} (\bibinfo {year} {2005})},\
  \Eprint {http://arxiv.org/abs/hep-th/0502189} {arXiv:hep-th/0502189 [hep-th]}
  \BibitemShut {NoStop}%
\bibitem [{\citenamefont {Babichev}\ \emph
  {et~al.}(2018{\natexlab{b}})\citenamefont {Babichev}, \citenamefont
  {Charmousis}, \citenamefont {Esposito-Far\`ese},\ and\ \citenamefont
  {Leh\'ebel}}]{Babichev:2018uiw}%
  \BibitemOpen
  \bibfield  {author} {\bibinfo {author} {\bibfnamefont {E.}~\bibnamefont
  {Babichev}}, \bibinfo {author} {\bibfnamefont {C.}~\bibnamefont
  {Charmousis}}, \bibinfo {author} {\bibfnamefont {G.}~\bibnamefont
  {Esposito-Far\`ese}},  and \bibinfo {author} {\bibfnamefont {A.}~\bibnamefont
  {Leh\'ebel}},\ }\href {\doibase 10.1103/PhysRevD.98.104050} {\bibfield
  {journal} {\bibinfo  {journal} {\emph {Phys. Rev.}}\ }\textbf {\bibinfo
  {volume} {D98}},\ \bibinfo {pages} {104050} (\bibinfo {year}
  {2018}{\natexlab{b}})},\ \Eprint {http://arxiv.org/abs/1803.11444}
  {arXiv:1803.11444 [gr-qc]} \BibitemShut {NoStop}%
\bibitem [{\citenamefont {de~Rham}\ and\ \citenamefont
  {Matas}(2016)}]{deRham:2016wji}%
  \BibitemOpen
  \bibfield  {author} {\bibinfo {author} {\bibfnamefont {C.}~\bibnamefont
  {de~Rham}} and \bibinfo {author} {\bibfnamefont {A.}~\bibnamefont {Matas}},\
  }\href {\doibase 10.1088/1475-7516/2016/06/041} {\bibfield  {journal}
  {\bibinfo  {journal} {\emph {JCAP}}\ }\textbf {\bibinfo {volume} {1606}},\
  \bibinfo {pages} {041} (\bibinfo {year} {2016})},\ \Eprint
  {http://arxiv.org/abs/1604.08638} {arXiv:1604.08638 [hep-th]} \BibitemShut
  {NoStop}%
\bibitem [{\citenamefont {Langlois}\ \emph {et~al.}(2017)\citenamefont
  {Langlois}, \citenamefont {Mancarella}, \citenamefont {Noui},\ and\
  \citenamefont {Vernizzi}}]{Langlois:2017mxy}%
  \BibitemOpen
  \bibfield  {author} {\bibinfo {author} {\bibfnamefont {D.}~\bibnamefont
  {Langlois}}, \bibinfo {author} {\bibfnamefont {M.}~\bibnamefont
  {Mancarella}}, \bibinfo {author} {\bibfnamefont {K.}~\bibnamefont {Noui}},
  and \bibinfo {author} {\bibfnamefont {F.}~\bibnamefont {Vernizzi}},\ }\href
  {\doibase 10.1088/1475-7516/2017/05/033} {\bibfield  {journal} {\bibinfo
  {journal} {\emph {JCAP}}\ }\textbf {\bibinfo {volume} {1705}},\ \bibinfo
  {pages} {033} (\bibinfo {year} {2017})},\ \Eprint
  {http://arxiv.org/abs/1703.03797} {arXiv:1703.03797 [hep-th]} \BibitemShut
  {NoStop}%
\bibitem [{\citenamefont {Bekenstein}(1993)}]{Bekenstein:1992pj}%
  \BibitemOpen
  \bibfield  {author} {\bibinfo {author} {\bibfnamefont {J.~D.}\ \bibnamefont
  {Bekenstein}},\ }\href {\doibase 10.1103/PhysRevD.48.3641} {\bibfield
  {journal} {\bibinfo  {journal} {\emph {Phys. Rev.}}\ }\textbf {\bibinfo
  {volume} {D48}},\ \bibinfo {pages} {3641} (\bibinfo {year} {1993})},\ \Eprint
  {http://arxiv.org/abs/gr-qc/9211017} {arXiv:gr-qc/9211017 [gr-qc]}
  \BibitemShut {NoStop}%
\bibitem [{\citenamefont {Zumalac\'arregui}\ and\ \citenamefont
  {Garc\'ia-Bellido}(2014)}]{Zumalacarregui:2013pma}%
  \BibitemOpen
  \bibfield  {author} {\bibinfo {author} {\bibfnamefont {M.}~\bibnamefont
  {Zumalac\'arregui}} and \bibinfo {author} {\bibfnamefont {J.}~\bibnamefont
  {Garc\'ia-Bellido}},\ }\href {\doibase 10.1103/PhysRevD.89.064046} {\bibfield
   {journal} {\bibinfo  {journal} {\emph {Phys. Rev.}}\ }\textbf {\bibinfo
  {volume} {D89}},\ \bibinfo {pages} {064046} (\bibinfo {year} {2014})},\
  \Eprint {http://arxiv.org/abs/1308.4685} {arXiv:1308.4685 [gr-qc]}
  \BibitemShut {NoStop}%
\bibitem [{\citenamefont {Takahashi}\ \emph {et~al.}(2017)\citenamefont
  {Takahashi}, \citenamefont {Motohashi}, \citenamefont {Suyama},\ and\
  \citenamefont {Kobayashi}}]{Takahashi:2017zgr}%
  \BibitemOpen
  \bibfield  {author} {\bibinfo {author} {\bibfnamefont {K.}~\bibnamefont
  {Takahashi}}, \bibinfo {author} {\bibfnamefont {H.}~\bibnamefont
  {Motohashi}}, \bibinfo {author} {\bibfnamefont {T.}~\bibnamefont {Suyama}},
  and \bibinfo {author} {\bibfnamefont {T.}~\bibnamefont {Kobayashi}},\ }\href
  {\doibase 10.1103/PhysRevD.95.084053} {\bibfield  {journal} {\bibinfo
  {journal} {\emph {Phys. Rev.}}\ }\textbf {\bibinfo {volume} {D95}},\ \bibinfo
  {pages} {084053} (\bibinfo {year} {2017})},\ \Eprint
  {http://arxiv.org/abs/1702.01849} {arXiv:1702.01849 [gr-qc]} \BibitemShut
  {NoStop}%
\bibitem [{\citenamefont {Dom\`enech}\ \emph {et~al.}(2015)\citenamefont
  {Dom\`enech}, \citenamefont {Mukohyama}, \citenamefont {Namba}, \citenamefont
  {Naruko}, \citenamefont {Saitou},\ and\ \citenamefont
  {Watanabe}}]{Domenech:2015tca}%
  \BibitemOpen
  \bibfield  {author} {\bibinfo {author} {\bibfnamefont {G.}~\bibnamefont
  {Dom\`enech}}, \bibinfo {author} {\bibfnamefont {S.}~\bibnamefont
  {Mukohyama}}, \bibinfo {author} {\bibfnamefont {R.}~\bibnamefont {Namba}},
  \bibinfo {author} {\bibfnamefont {A.}~\bibnamefont {Naruko}}, \bibinfo
  {author} {\bibfnamefont {R.}~\bibnamefont {Saitou}},  and \bibinfo {author}
  {\bibfnamefont {Y.}~\bibnamefont {Watanabe}},\ }\href {\doibase
  10.1103/PhysRevD.92.084027} {\bibfield  {journal} {\bibinfo  {journal} {\emph
  {Phys. Rev.}}\ }\textbf {\bibinfo {volume} {D92}},\ \bibinfo {pages} {084027}
  (\bibinfo {year} {2015})},\ \Eprint {http://arxiv.org/abs/1507.05390}
  {arXiv:1507.05390 [hep-th]} \BibitemShut {NoStop}%
\bibitem [{\citenamefont {Abbott}\ \emph
  {et~al.}(2017{\natexlab{a}})\citenamefont {Abbott} \emph
  {et~al.}}]{TheLIGOScientific:2017qsa}%
  \BibitemOpen
  \bibfield  {author} {\bibinfo {author} {\bibfnamefont {B.}~\bibnamefont
  {Abbott}} \emph {et~al.} (\bibinfo {collaboration} {LIGO Scientific,
  Virgo}),\ }\href {\doibase 10.1103/PhysRevLett.119.161101} {\bibfield
  {journal} {\bibinfo  {journal} {\emph {Phys. Rev. Lett.}}\ }\textbf {\bibinfo
  {volume} {119}},\ \bibinfo {pages} {161101} (\bibinfo {year}
  {2017}{\natexlab{a}})},\ \Eprint {http://arxiv.org/abs/1710.05832}
  {arXiv:1710.05832 [gr-qc]} \BibitemShut {NoStop}%
\bibitem [{\citenamefont {Abbott}\ \emph
  {et~al.}(2017{\natexlab{b}})\citenamefont {Abbott} \emph
  {et~al.}}]{GBM:2017lvd}%
  \BibitemOpen
  \bibfield  {author} {\bibinfo {author} {\bibfnamefont {B.~P.}\ \bibnamefont
  {Abbott}} \emph {et~al.} (\bibinfo {collaboration} {LIGO Scientific, Virgo,
  Fermi GBM, INTEGRAL, IceCube, AstroSat Cadmium Zinc Telluride Imager Team,
  IPN, Insight-Hxmt, ANTARES, Swift, AGILE Team, 1M2H Team, Dark Energy Camera
  GW-EM, DES, DLT40, GRAWITA, Fermi-LAT, ATCA, ASKAP, Las Cumbres Observatory
  Group, OzGrav, DWF (Deeper Wider Faster Program), AST3, CAASTRO, VINROUGE,
  MASTER, J-GEM, GROWTH, JAGWAR, CaltechNRAO, TTU-NRAO, NuSTAR, Pan-STARRS,
  MAXI Team, TZAC Consortium, KU, Nordic Optical Telescope, ePESSTO, GROND,
  Texas Tech University, SALT Group, TOROS, BOOTES, MWA, CALET, IKI-GW
  Follow-up, H.E.S.S., LOFAR, LWA, HAWC, Pierre Auger, ALMA, Euro VLBI Team, Pi
  of Sky, Chandra Team at McGill University, DFN, ATLAS Telescopes, High Time
  Resolution Universe Survey, RIMAS, RATIR, SKA South Africa/MeerKAT}),\ }\href
  {\doibase 10.3847/2041-8213/aa91c9} {\bibfield  {journal} {\bibinfo
  {journal} {\emph {Astrophys. J.}}\ }\textbf {\bibinfo {volume} {848}},\
  \bibinfo {pages} {L12} (\bibinfo {year} {2017}{\natexlab{b}})},\ \Eprint
  {http://arxiv.org/abs/1710.05833} {arXiv:1710.05833 [astro-ph.HE]}
  \BibitemShut {NoStop}%
\bibitem [{\citenamefont {Abbott}\ \emph
  {et~al.}(2017{\natexlab{c}})\citenamefont {Abbott} \emph
  {et~al.}}]{Monitor:2017mdv}%
  \BibitemOpen
  \bibfield  {author} {\bibinfo {author} {\bibfnamefont {B.~P.}\ \bibnamefont
  {Abbott}} \emph {et~al.} (\bibinfo {collaboration} {LIGO Scientific, Virgo,
  Fermi-GBM, INTEGRAL}),\ }\href {\doibase 10.3847/2041-8213/aa920c} {\bibfield
   {journal} {\bibinfo  {journal} {\emph {Astrophys. J.}}\ }\textbf {\bibinfo
  {volume} {848}},\ \bibinfo {pages} {L13} (\bibinfo {year}
  {2017}{\natexlab{c}})},\ \Eprint {http://arxiv.org/abs/1710.05834}
  {arXiv:1710.05834 [astro-ph.HE]} \BibitemShut {NoStop}%
\bibitem [{\citenamefont {Langlois}\ \emph {et~al.}(2018)\citenamefont
  {Langlois}, \citenamefont {Saito}, \citenamefont {Yamauchi},\ and\
  \citenamefont {Noui}}]{Langlois:2017dyl}%
  \BibitemOpen
  \bibfield  {author} {\bibinfo {author} {\bibfnamefont {D.}~\bibnamefont
  {Langlois}}, \bibinfo {author} {\bibfnamefont {R.}~\bibnamefont {Saito}},
  \bibinfo {author} {\bibfnamefont {D.}~\bibnamefont {Yamauchi}},  and \bibinfo
  {author} {\bibfnamefont {K.}~\bibnamefont {Noui}},\ }\href {\doibase
  10.1103/PhysRevD.97.061501} {\bibfield  {journal} {\bibinfo  {journal} {\emph
  {Phys. Rev.}}\ }\textbf {\bibinfo {volume} {D97}},\ \bibinfo {pages} {061501}
  (\bibinfo {year} {2018})},\ \Eprint {http://arxiv.org/abs/1711.07403}
  {arXiv:1711.07403 [gr-qc]} \BibitemShut {NoStop}%
\bibitem [{\citenamefont {Creminelli}\ \emph {et~al.}(2018)\citenamefont
  {Creminelli}, \citenamefont {Lewandowski}, \citenamefont {Tambalo},\ and\
  \citenamefont {Vernizzi}}]{Creminelli:2018xsv}%
  \BibitemOpen
  \bibfield  {author} {\bibinfo {author} {\bibfnamefont {P.}~\bibnamefont
  {Creminelli}}, \bibinfo {author} {\bibfnamefont {M.}~\bibnamefont
  {Lewandowski}}, \bibinfo {author} {\bibfnamefont {G.}~\bibnamefont
  {Tambalo}},  and \bibinfo {author} {\bibfnamefont {F.}~\bibnamefont
  {Vernizzi}},\ }\href {\doibase 10.1088/1475-7516/2018/12/025} {\bibfield
  {journal} {\bibinfo  {journal} {\emph {JCAP}}\ }\textbf {\bibinfo {volume}
  {1812}},\ \bibinfo {pages} {025} (\bibinfo {year} {2018})},\ \Eprint
  {http://arxiv.org/abs/1809.03484} {arXiv:1809.03484 [astro-ph.CO]}
  \BibitemShut {NoStop}%
\bibitem [{\citenamefont {de~Rham}\ and\ \citenamefont
  {Melville}(2018)}]{deRham:2018red}%
  \BibitemOpen
  \bibfield  {author} {\bibinfo {author} {\bibfnamefont {C.}~\bibnamefont
  {de~Rham}} and \bibinfo {author} {\bibfnamefont {S.}~\bibnamefont
  {Melville}},\ }\href {\doibase 10.1103/PhysRevLett.121.221101} {\bibfield
  {journal} {\bibinfo  {journal} {\emph {Phys. Rev. Lett.}}\ }\textbf {\bibinfo
  {volume} {121}},\ \bibinfo {pages} {221101} (\bibinfo {year} {2018})},\
  \Eprint {http://arxiv.org/abs/1806.09417} {arXiv:1806.09417 [hep-th]}
  \BibitemShut {NoStop}%
\bibitem [{\citenamefont {Kobayashi}\ and\ \citenamefont
  {Hiramatsu}(2018)}]{Kobayashi:2018xvr}%
  \BibitemOpen
  \bibfield  {author} {\bibinfo {author} {\bibfnamefont {T.}~\bibnamefont
  {Kobayashi}} and \bibinfo {author} {\bibfnamefont {T.}~\bibnamefont
  {Hiramatsu}},\ }\href {\doibase 10.1103/PhysRevD.97.104012} {\bibfield
  {journal} {\bibinfo  {journal} {\emph {Phys. Rev.}}\ }\textbf {\bibinfo
  {volume} {D97}},\ \bibinfo {pages} {104012} (\bibinfo {year} {2018})},\
  \Eprint {http://arxiv.org/abs/1803.10510} {arXiv:1803.10510 [gr-qc]}
  \BibitemShut {NoStop}%
\bibitem [{\citenamefont {Afshordi}\ \emph {et~al.}(2007)\citenamefont
  {Afshordi}, \citenamefont {Chung},\ and\ \citenamefont
  {Geshnizjani}}]{Afshordi:2006ad}%
  \BibitemOpen
  \bibfield  {author} {\bibinfo {author} {\bibfnamefont {N.}~\bibnamefont
  {Afshordi}}, \bibinfo {author} {\bibfnamefont {D.~J.~H.}\ \bibnamefont
  {Chung}},  and \bibinfo {author} {\bibfnamefont {G.}~\bibnamefont
  {Geshnizjani}},\ }\href {\doibase 10.1103/PhysRevD.75.083513} {\bibfield
  {journal} {\bibinfo  {journal} {\emph {Phys. Rev.}}\ }\textbf {\bibinfo
  {volume} {D75}},\ \bibinfo {pages} {083513} (\bibinfo {year} {2007})},\
  \Eprint {http://arxiv.org/abs/hep-th/0609150} {arXiv:hep-th/0609150 [hep-th]}
  \BibitemShut {NoStop}%
\bibitem [{\citenamefont {Iyonaga}\ \emph {et~al.}(2018)\citenamefont
  {Iyonaga}, \citenamefont {Takahashi},\ and\ \citenamefont
  {Kobayashi}}]{Iyonaga:2018vnu}%
  \BibitemOpen
  \bibfield  {author} {\bibinfo {author} {\bibfnamefont {A.}~\bibnamefont
  {Iyonaga}}, \bibinfo {author} {\bibfnamefont {K.}~\bibnamefont {Takahashi}},
  and \bibinfo {author} {\bibfnamefont {T.}~\bibnamefont {Kobayashi}},\ }\href
  {\doibase 10.1088/1475-7516/2018/12/002} {\bibfield  {journal} {\bibinfo
  {journal} {\emph {JCAP}}\ }\textbf {\bibinfo {volume} {1812}},\ \bibinfo
  {pages} {002} (\bibinfo {year} {2018})},\ \Eprint
  {http://arxiv.org/abs/1809.10935} {arXiv:1809.10935 [gr-qc]} \BibitemShut
  {NoStop}%
\bibitem [{\citenamefont {de~Rham}\ and\ \citenamefont
  {Motohashi}(2017)}]{deRham:2016ged}%
  \BibitemOpen
  \bibfield  {author} {\bibinfo {author} {\bibfnamefont {C.}~\bibnamefont
  {de~Rham}} and \bibinfo {author} {\bibfnamefont {H.}~\bibnamefont
  {Motohashi}},\ }\href {\doibase 10.1103/PhysRevD.95.064008} {\bibfield
  {journal} {\bibinfo  {journal} {\emph {Phys. Rev.}}\ }\textbf {\bibinfo
  {volume} {D95}},\ \bibinfo {pages} {064008} (\bibinfo {year} {2017})},\
  \Eprint {http://arxiv.org/abs/1611.05038} {arXiv:1611.05038 [hep-th]}
  \BibitemShut {NoStop}%
\bibitem [{\citenamefont {Motohashi}\ and\ \citenamefont
  {Minamitsuji}(2018)}]{Motohashi:2018wdq}%
  \BibitemOpen
  \bibfield  {author} {\bibinfo {author} {\bibfnamefont {H.}~\bibnamefont
  {Motohashi}} and \bibinfo {author} {\bibfnamefont {M.}~\bibnamefont
  {Minamitsuji}},\ }\href {\doibase 10.1016/j.physletb.2018.04.041} {\bibfield
  {journal} {\bibinfo  {journal} {\emph {Phys. Lett.}}\ }\textbf {\bibinfo
  {volume} {B781}},\ \bibinfo {pages} {728} (\bibinfo {year} {2018})},\ \Eprint
  {http://arxiv.org/abs/1804.01731} {arXiv:1804.01731 [gr-qc]} \BibitemShut
  {NoStop}%
\bibitem [{\citenamefont {Babichev}\ \emph {et~al.}(2017)\citenamefont
  {Babichev}, \citenamefont {Charmousis},\ and\ \citenamefont
  {Leh\'ebel}}]{Babichev:2017guv}%
  \BibitemOpen
  \bibfield  {author} {\bibinfo {author} {\bibfnamefont {E.}~\bibnamefont
  {Babichev}}, \bibinfo {author} {\bibfnamefont {C.}~\bibnamefont
  {Charmousis}},  and \bibinfo {author} {\bibfnamefont {A.}~\bibnamefont
  {Leh\'ebel}},\ }\href {\doibase 10.1088/1475-7516/2017/04/027} {\bibfield
  {journal} {\bibinfo  {journal} {\emph {JCAP}}\ }\textbf {\bibinfo {volume}
  {1704}},\ \bibinfo {pages} {027} (\bibinfo {year} {2017})},\ \Eprint
  {http://arxiv.org/abs/1702.01938} {arXiv:1702.01938 [gr-qc]} \BibitemShut
  {NoStop}%
\bibitem [{\citenamefont {Motohashi}\ and\ \citenamefont
  {Suyama}(2011)}]{Motohashi:2011pw}%
  \BibitemOpen
  \bibfield  {author} {\bibinfo {author} {\bibfnamefont {H.}~\bibnamefont
  {Motohashi}} and \bibinfo {author} {\bibfnamefont {T.}~\bibnamefont
  {Suyama}},\ }\href {\doibase 10.1103/PhysRevD.84.084041} {\bibfield
  {journal} {\bibinfo  {journal} {\emph {Phys. Rev.}}\ }\textbf {\bibinfo
  {volume} {D84}},\ \bibinfo {pages} {084041} (\bibinfo {year} {2011})},\
  \Eprint {http://arxiv.org/abs/1107.3705} {arXiv:1107.3705 [gr-qc]}
  \BibitemShut {NoStop}%
\bibitem [{\citenamefont {Motohashi}\ and\ \citenamefont
  {Suyama}(2012)}]{Motohashi:2011ds}%
  \BibitemOpen
  \bibfield  {author} {\bibinfo {author} {\bibfnamefont {H.}~\bibnamefont
  {Motohashi}} and \bibinfo {author} {\bibfnamefont {T.}~\bibnamefont
  {Suyama}},\ }\href {\doibase 10.1103/PhysRevD.85.044054} {\bibfield
  {journal} {\bibinfo  {journal} {\emph {Phys. Rev.}}\ }\textbf {\bibinfo
  {volume} {D85}},\ \bibinfo {pages} {044054} (\bibinfo {year} {2012})},\
  \Eprint {http://arxiv.org/abs/1110.6241} {arXiv:1110.6241 [gr-qc]}
  \BibitemShut {NoStop}%
\bibitem [{\citenamefont {Regge}\ and\ \citenamefont
  {Wheeler}(1957)}]{Regge:1957td}%
  \BibitemOpen
  \bibfield  {author} {\bibinfo {author} {\bibfnamefont {T.}~\bibnamefont
  {Regge}} and \bibinfo {author} {\bibfnamefont {J.~A.}\ \bibnamefont
  {Wheeler}},\ }\href {\doibase 10.1103/PhysRev.108.1063} {\bibfield  {journal}
  {\bibinfo  {journal} {\emph {Phys. Rev.}}\ }\textbf {\bibinfo {volume}
  {108}},\ \bibinfo {pages} {1063} (\bibinfo {year} {1957})}\BibitemShut
  {NoStop}%
\bibitem [{\citenamefont {Motohashi}\ \emph
  {et~al.}(2016{\natexlab{b}})\citenamefont {Motohashi}, \citenamefont
  {Suyama},\ and\ \citenamefont {Takahashi}}]{Motohashi:2016prk}%
  \BibitemOpen
  \bibfield  {author} {\bibinfo {author} {\bibfnamefont {H.}~\bibnamefont
  {Motohashi}}, \bibinfo {author} {\bibfnamefont {T.}~\bibnamefont {Suyama}},
  and \bibinfo {author} {\bibfnamefont {K.}~\bibnamefont {Takahashi}},\ }\href
  {\doibase 10.1103/PhysRevD.94.124021} {\bibfield  {journal} {\bibinfo
  {journal} {\emph {Phys. Rev.}}\ }\textbf {\bibinfo {volume} {D94}},\ \bibinfo
  {pages} {124021} (\bibinfo {year} {2016}{\natexlab{b}})},\ \Eprint
  {http://arxiv.org/abs/1608.00071} {arXiv:1608.00071 [gr-qc]} \BibitemShut
  {NoStop}%
\bibitem [{\citenamefont {De~Felice}\ \emph {et~al.}(2011)\citenamefont
  {De~Felice}, \citenamefont {Suyama},\ and\ \citenamefont
  {Tanaka}}]{DeFelice:2011ka}%
  \BibitemOpen
  \bibfield  {author} {\bibinfo {author} {\bibfnamefont {A.}~\bibnamefont
  {De~Felice}}, \bibinfo {author} {\bibfnamefont {T.}~\bibnamefont {Suyama}},
  and \bibinfo {author} {\bibfnamefont {T.}~\bibnamefont {Tanaka}},\ }\href
  {\doibase 10.1103/PhysRevD.83.104035} {\bibfield  {journal} {\bibinfo
  {journal} {\emph {Phys. Rev.}}\ }\textbf {\bibinfo {volume} {D83}},\ \bibinfo
  {pages} {104035} (\bibinfo {year} {2011})},\ \Eprint
  {http://arxiv.org/abs/1102.1521} {arXiv:1102.1521 [gr-qc]} \BibitemShut
  {NoStop}%
\bibitem [{\citenamefont {Kobayashi}\ \emph {et~al.}(2012)\citenamefont
  {Kobayashi}, \citenamefont {Motohashi},\ and\ \citenamefont
  {Suyama}}]{Kobayashi:2012kh}%
  \BibitemOpen
  \bibfield  {author} {\bibinfo {author} {\bibfnamefont {T.}~\bibnamefont
  {Kobayashi}}, \bibinfo {author} {\bibfnamefont {H.}~\bibnamefont
  {Motohashi}},  and \bibinfo {author} {\bibfnamefont {T.}~\bibnamefont
  {Suyama}},\ }\href {\doibase 10.1103/PhysRevD.85.084025} {\bibfield
  {journal} {\bibinfo  {journal} {\emph {Phys. Rev.}}\ }\textbf {\bibinfo
  {volume} {D85}},\ \bibinfo {pages} {084025} (\bibinfo {year} {2012})},\
  \bibinfo {note} {[Erratum:
  \href{https://doi.org/10.1103/PhysRevD.96.109903}{Phys. Rev. {\bf D96},
  no.10, 109903 (2017)}]},\ \Eprint {http://arxiv.org/abs/1202.4893}
  {arXiv:1202.4893 [gr-qc]} \BibitemShut {NoStop}%
\bibitem [{\citenamefont {Takahashi}\ \emph {et~al.}(2016)\citenamefont
  {Takahashi}, \citenamefont {Suyama},\ and\ \citenamefont
  {Kobayashi}}]{Takahashi:2015pad}%
  \BibitemOpen
  \bibfield  {author} {\bibinfo {author} {\bibfnamefont {K.}~\bibnamefont
  {Takahashi}}, \bibinfo {author} {\bibfnamefont {T.}~\bibnamefont {Suyama}},
  and \bibinfo {author} {\bibfnamefont {T.}~\bibnamefont {Kobayashi}},\ }\href
  {\doibase 10.1103/PhysRevD.93.064068} {\bibfield  {journal} {\bibinfo
  {journal} {\emph {Phys. Rev.}}\ }\textbf {\bibinfo {volume} {D93}},\ \bibinfo
  {pages} {064068} (\bibinfo {year} {2016})},\ \Eprint
  {http://arxiv.org/abs/1511.06083} {arXiv:1511.06083 [gr-qc]} \BibitemShut
  {NoStop}%
\bibitem [{\citenamefont {Tretyakova}\ and\ \citenamefont
  {Takahashi}(2017)}]{Tretyakova:2017lyg}%
  \BibitemOpen
  \bibfield  {author} {\bibinfo {author} {\bibfnamefont {D.~A.}\ \bibnamefont
  {Tretyakova}} and \bibinfo {author} {\bibfnamefont {K.}~\bibnamefont
  {Takahashi}},\ }\href {\doibase 10.1088/1361-6382/aa8057} {\bibfield
  {journal} {\bibinfo  {journal} {\emph {Class. Quant. Grav.}}\ }\textbf
  {\bibinfo {volume} {34}},\ \bibinfo {pages} {175007} (\bibinfo {year}
  {2017})},\ \Eprint {http://arxiv.org/abs/1702.03502} {arXiv:1702.03502
  [gr-qc]} \BibitemShut {NoStop}%
\bibitem [{\citenamefont {Kodama}\ and\ \citenamefont
  {Ishibashi}(2003)}]{Kodama:2003jz}%
  \BibitemOpen
  \bibfield  {author} {\bibinfo {author} {\bibfnamefont {H.}~\bibnamefont
  {Kodama}} and \bibinfo {author} {\bibfnamefont {A.}~\bibnamefont
  {Ishibashi}},\ }\href {\doibase 10.1143/PTP.110.701} {\bibfield  {journal}
  {\bibinfo  {journal} {\emph {Prog. Theor. Phys.}}\ }\textbf {\bibinfo
  {volume} {110}},\ \bibinfo {pages} {701} (\bibinfo {year} {2003})},\ \Eprint
  {http://arxiv.org/abs/hep-th/0305147} {arXiv:hep-th/0305147 [hep-th]}
  \BibitemShut {NoStop}%
\bibitem [{\citenamefont {Ishibashi}\ and\ \citenamefont
  {Kodama}(2003)}]{Ishibashi:2003ap}%
  \BibitemOpen
  \bibfield  {author} {\bibinfo {author} {\bibfnamefont {A.}~\bibnamefont
  {Ishibashi}} and \bibinfo {author} {\bibfnamefont {H.}~\bibnamefont
  {Kodama}},\ }\href {\doibase 10.1143/PTP.110.901} {\bibfield  {journal}
  {\bibinfo  {journal} {\emph {Prog. Theor. Phys.}}\ }\textbf {\bibinfo
  {volume} {110}},\ \bibinfo {pages} {901} (\bibinfo {year} {2003})},\ \Eprint
  {http://arxiv.org/abs/hep-th/0305185} {arXiv:hep-th/0305185 [hep-th]}
  \BibitemShut {NoStop}%
\bibitem [{\citenamefont {Kodama}\ and\ \citenamefont
  {Ishibashi}(2004)}]{Kodama:2003kk}%
  \BibitemOpen
  \bibfield  {author} {\bibinfo {author} {\bibfnamefont {H.}~\bibnamefont
  {Kodama}} and \bibinfo {author} {\bibfnamefont {A.}~\bibnamefont
  {Ishibashi}},\ }\href {\doibase 10.1143/PTP.111.29} {\bibfield  {journal}
  {\bibinfo  {journal} {\emph {Prog. Theor. Phys.}}\ }\textbf {\bibinfo
  {volume} {111}},\ \bibinfo {pages} {29} (\bibinfo {year} {2004})},\ \Eprint
  {http://arxiv.org/abs/hep-th/0308128} {arXiv:hep-th/0308128 [hep-th]}
  \BibitemShut {NoStop}%
\end{thebibliography}%

\end{document}